\documentclass[12pt]{article} 
\usepackage{graphicx}
\usepackage{amsmath}
\usepackage{amssymb}

 \newcommand{\R}{{\mathbb R}}
 \newcommand{\Espe}{{\mathbb E}}
 \newcommand{\CQFD}
 {
 \mbox{}
 \nolinebreak
 \hfill
 \rule{2mm}{2mm}
 \medbreak
 \par
 }

 \newtheorem{Theo}{Theorem}

 \newtheorem{lemma}{Lemma}
 \newtheorem{Coro}{Corollary}
 \newtheorem{Remark}{Remark}

\begin{document}

\pagestyle{empty}
\font\geant=cmssbx10 at 16pt
\font \nain=cmss10 at 12pt
\begin{center}

\geant LINEAR PROGRAMMING PROBLEMS
\vskip 10pt
FOR $L_1$- OPTIMAL FRONTIER ESTIMATION
\vskip 18pt
\nain  S. GIRARD, A. IOUDITSKI and  A. NAZIN\\
\vskip 20pt
\vskip 20pt

\end{center}

\noindent {\bf Abstract}\\

\noindent We propose new \textit{optimal} estimators for the
Lipschitz frontier of a set of points.
They are defined as kernel estimators being sufficiently regular, covering
all the points and whose associated support is of smallest surface. The
estimators are written as linear combinations of kernel functions applied to the
points of the sample. The coefficients of the linear combination are then
computed by solving related linear programming problem.
The $L_1$ error between the estimated and the true frontier function
with a known Lipschitz constant is
shown to be almost surely converging to zero, and the rate of convergence
is proved to be optimal. \\

\noindent {\bf Contact information}\\

\noindent St\'ephane Girard and Anatoli Iouditski: SMS/LMC, Universit\'e Grenoble I, BP 53,\\
38041 Grenoble cedex 9, France.
{\tt \{Stephane.Girard, Anatoli.Iouditski\}@imag.fr}  \\

\noindent Alexander Nazin: Institute of Control Sciences RAS, Profsoyuznaya str., 65, \\
117997 Moscow, Russia.  {\tt nazine@ipu.rssi.ru}\\

\noindent {\bf Acknowledgements}\\

\noindent Financial support from the IAP research network nr P5/24 of the
Belgian Government (Federal Office for Scientific, Technical and Cultural Affairs) is
gratefully acknowledged.\\

\noindent The work of A.~Nazin has been carried out during his stay in
 SMS/LMC, Universit\'e Grenoble-I, and INRIA Rh\^one-Alpes
hold in June and October--November 2004.

\newpage

\section{Introduction}
\label{secIntro}

Many proposals are given in the literature for estimating a set $S$ 
given a finite random set of points drawn from the interior.
This problem of edge or support estimation arises in classification
({\sc Hardy} \& {\sc Rasson}~\cite{HarRas}),
clustering problems ({\sc Hartigan}~\cite{Hart}),
discriminant analysis ({\sc Baufays} \& {\sc Rasson}~\cite{BauRas}),
and outliers detection ({\sc Devroye} \& {\sc Wise}~\cite{DevWis}).  
Applications are found in medical diagnosis ({\sc Tarassenko}
{\it et al}~\cite{THCB}) as well as in condition monitoring of 
machines ({\sc Devroye} \& {\sc Wise}~\cite{DevWis}).
In image analysis,
the segmentation problem can be considered under the support estimation
point of view, where the support is a convex bounded set in $\R^2$
({\sc Korostelev} \& {\sc Tsybakov}~\cite{KorTsy2}).
We also point out some applications in econometrics
(e.g. {\sc Deprins}, {\it et al}~\cite{DepSimTul}).
In such cases, the unknown support can be written
\begin{equation}
\label{defS}
S\triangleq\{(x,y):\;~0\leq~x\leq~1\;~;~\;~0\leq~y~\leq~f(x)\},
\end{equation}
where $f$ is an unknown function. 
Here, the problem reduces to estimating $f$, called the production frontier
(see for instance {\sc H\"ardle} {\it et al}~\cite{Hardle}).
The data consist of pair $(X,Y)$
where $X$ represents the input (labor, energy or capital) used to produce
an output $Y$ in a given firm. In such a framework, the value $f(x)$
 can be interpreted as the maximum level of output which is attainable
for the level of input $x$.\\

\noindent 
An early paper was written by {\sc Geffroy}~\cite{Geff1} for 
independent identically distributed observations
from a density $\phi$.
The proposed estimator is a kind of histogram based on the extreme values of the sample.
This work was extended in two main directions.\\

\noindent On the one hand, piecewise polynomials estimators were introduced.  
They are  defined locally on a given slice as the lowest
polynomial of fixed degree covering all the points in the considered slice.
Their optimality in an asymptotic minimax sense is proved under weak
assumptions on the rate of decrease $\alpha$ of the density $\phi$ towards 0
by {\sc Korostelev} \& {\sc Tsybakov}~\cite{KorTsy2} and by
{\sc H\"ardle} {\it et al}~\cite{Hardle2}. Extreme values methods
are then proposed by {\sc Hall} {\it et al}~\cite{Hall} and by
{\sc Gijbels} \& {\sc Peng}~\cite{Gijbels} to estimate the parameter $\alpha$.
Estimating $f$ can also been considered as a regression problem 
$Y=f(X)+\varepsilon$ with negative noise $\varepsilon$. 
In this context,
local polynomial estimates are introduced, see {\sc Knight}~\cite{KK}, or
{\sc Hall} {\it et al}~\cite{Hall2} for a similar approach.\\

\noindent
On the other hand, different propositions for smoothing
Geffroy's estimator were made in the case of a Poisson point
process. {\sc Girard} \& {\sc Jacob}~\cite{GirJac3}
 introduced estimators based on kernel regressions
and orthogonal series method~\cite{GirJac,GirJac2}.
In the same spirit, {\sc Gardes}~\cite{Gardes} proposed a Faber-Shauder estimator. 
{\sc Girard} \& {\sc Menneteau}~\cite{GirMen} introduced a general framework
for studying estimators of this type and generalized them to
supports writing 
$$
S=\{(x,y):\;~x\in E\;~;~\;~0\leq~y~\leq~f(x)\},
$$
where $f$ is an unknown function and $E$ an arbitrary set. 
In each case, the limit distribution of the estimator is established. 
 We also refer to {\sc Abbar}~\cite{Abb} and
 {\sc Jacob} \& {\sc Suquet}~\cite{JacSuq} who used a similar
smoothing approach, although their estimators are not based  on
the extreme values of the Poisson process.\\

\noindent 
The estimator proposed in {\sc Bouchard} {\it et al}~\cite{BGIN} can
be considered to belong to the intersect of these two directions.  
From the practical point of view, 
it is defined as a kernel estimator obtained by smoothing
some selected points of the sample.   These points are
chosen automatically by solving a linear programming problem
to obtain an estimated
support covering all the points and with smallest surface.
From the theoretical point of view, this estimator is shown
to be consistent for the $L_1$ norm. \\

\noindent 
In this paper, we propose several modifications of the above method.  
First, a bias corrected kernel is proposed.   Second,
some regularity constraints are introduced in the optimization
problem.   We show that the resulting estimator reaches the
optimal minimax $L_1$ rate (up to a logarithmic factor).
The estimator is defined in Section~\ref{subsecLPprobl}. 
Some preliminary properties are established in Section~\ref{preliresult},
and the main result is presented in Section~\ref{secmain}.
Proofs are postponed to Section~\ref{proofs}.

\section{Boundary estimator}
\label{subsecLPprobl}

Let all the random variables be defined on a probability space
$(\Omega,\mathcal{F},P)$.
The problem under consideration is to estimate an unknown positive
function $f: [0,1]\to (0,\infty)$ on the basis of observations
$(X_i,Y_i)_{i=1,\dots,N}$
with independent pairs
  $(X_i,Y_i)$ being uniformly distributed in the set $S$
  defined as 
  \begin{equation}\label{defS1}
  S \triangleq \{(x,y) : \, 0  \leq x\leq 1\,,\,\, 0\leq y\leq f(x) \}\,.
 \end{equation}
Letting
\begin{equation}\label{eq:Cf}
    C_f \triangleq \displaystyle\int_0^1 f(u)\,du 
    \,,
\end{equation}
each variable $X_i$ is distributed in $[0,1]$ with p.d.f.
$f(\cdot)/C_f$
while $Y_i$ has the uniform conditional distribution with respect to $X_i$
in the interval $[0,f(X_i)]$.
In what follows we assume $f\in\Sigma_{[0,1]}(\beta,L^{}_{f,\beta}\,)$, $0<\beta\leq 1$
that is function $f:[0,1]\to(0,\infty)$ is $\beta$-Lipschitz with constant $L^{}_{f,\beta}\,$\, :
\begin{equation}\label{LipschitzF}
|f(x)-f(u)|\leq L^{}_{f,\beta}\,|x-u|^\beta\quad\forall x,u\in [0,1]\,.
\end{equation}
\noindent The considered estimator $\widehat{f}_N:[0,1]\to[0,\infty)$
of the frontier is chosen from the family of
functions:
\noindent
\begin{equation}\label{estimator}
\left\{
\begin{array}{l}
\widehat{f}_N(x) = \displaystyle \sum_{i=1}^N \alpha_i\, K_h(x,X_i)\,,\qquad
\,K_h(x,t)=  \frac{g(x)}{h}\, K\left(\frac{x-t}{h}\right),\\
\alpha_i \geq 0,\qquad i=1,\dots,N,
\end{array}
\right.
\end{equation}
where $K$ is a sufficiently smooth
basic kernel function\, $K:\R\to[0,\infty)$ 
integrating to one and having the interval $[-1,1]$ as its support;
the bandwidth parameter $h\in(0,1/2)$ 
depends on $N$ such that $h\to 0$ as $N\to\infty$;
and the function 
\begin{equation}\label{Gcorrector}
g(x) = \left( \int_{(x-1)/h}^{x/h} K(t)\, dt \right)^{-1},
\quad x\in [0,1]\,,
\end{equation}
corrects the basic kernel $K$ at the boundaries, i.e., when $x\in[0,h)$ or $x\in(1-h,1]$.
Indeed, $g(x)\equiv 1$ on $x\in[h,1-h]$, while $g(x)>1$ when $x\in[0,h)$ or $x\in(1-h,1]$.

\noindent One may easily observe that
 \begin{equation}\label{IntCorKern1}
\int_0^1 K_h(x,u)\,du =1\qquad \forall\, x\in[0,1]
\end{equation}
and, consequently, due to interplacing the integral and the derivative,
\begin{equation}\label{IntCorKernDerx1}
\int_0^1 \frac{\partial}{\partial{x}} K_h(x,u)\;du =0\qquad \forall\, x\in[0,1]\,.
\end{equation}
Note, that equation (\ref{IntCorKernDerx1}) may be verified directly, 
as it is demonstrated in the Appendix, Subsection~\ref{subsecCorKernel}.

\noindent Denote $K_{\max}\triangleq\max K(t)$, $g_{\max}\triangleq\max g(x)$, as well as functionals
\begin{eqnarray}\label{Cbeta}
C_{\beta}(\varphi) &\triangleq& \displaystyle \int_{-1}^1 |t|^{\beta} |\varphi(t)|\, dt 
\,,\quad \varphi\in C^{\,0}([-1,1])\,,
\\ \label{Cbeta2}
C_{\beta}(K,K') &\triangleq& g_{\max}\,K_{\max}C_{\beta}(K) + C_{\beta}(K')\,.
\end{eqnarray}
We also denote by $L_{\varphi}$ a Lipschitz constant for function $\varphi :\R\to\R$, that is
\begin{equation}
|\varphi(s)-\varphi(t)|\leq L_{\varphi} |s-t| \mbox{ with }
L_{\varphi}<\infty. 
\end{equation}
The indicator function is denoted by $\mathbf{1}\{\cdot\}$ which equals $1$ if the argument condition
holds true, and $0$ otherwise.

\noindent As it is proved below in Lemma \ref{LemSurfApprox} the surface of the estimated support
  \begin{equation}\label{defSesti}
  \widehat S_N \triangleq \{(x,y) : \, 0  \leq x\leq 1\,,\,\, 0\leq y\leq \widehat f_N(x) \}
 \end{equation}
may be approximated as follows:
\begin{equation}\label{SurfApprox}
\int_0^1 \widehat{f}_N(x)\, dx = \sum_{i=1}^N \alpha_i +O(h) \,.
\end{equation}
This suggests to define the parameter vector
$\alpha=(\alpha_1,\dots,\alpha_N)^T$ as a solution to
the following optimization problem
\begin{eqnarray}\label{IPgoalS}
J_P^* &\triangleq& \min_\alpha \sum_{i=1}^N \alpha_i 
\\
\mathrm{subject~to} &&\nonumber
\\
&& \widehat{f}_N(X_i) \geq Y_i\,, \quad 
i=1,\dots,N\,,
\label{constr1S}\\
&& |\widehat{f}^{\;\prime}_N(X_i)|
 \leq L^{}_{f,\beta}\,\, g_{\max} C_{\beta}(K,K') \frac{\log N}{Nh^2}\,, 
\quad 
i=0,\dots,N+1\,,
 \label{constrPrimeS}\\ 
&&\sum_{i=1}^N \alpha_i \, \mathbf{1}\{ (j-1)/m_h\leq X_i < j/m_h\} \leq C_\alpha h
\,,\quad j=1,\dots,m_h \,,
 \label{constrSumAlp} \\
&&0\leq\alpha_i
\,,\quad i=1,\dots,N, \label{constr2S}
\end{eqnarray}
where parameter $m_h$ is defined to be the integer part of $1/h$\,.
This optimization problem may be formally written as linear program (LP)
\begin{eqnarray}\label{IPgoal}
J_P^* &\triangleq &\min_\alpha \,\mathbf{1}_N^T \alpha
\\
\mathrm{subject~to\phantom{aaa}} &&\nonumber
\\ 
 Y&\leq& A\alpha \,, \label{constr1}\\
 - L^{}_{f,\beta}\,\, g_{\max} C_{\beta}(K,K')\, 
 \frac{\log N}{Nh^2} \, \mathbf{1}_{N}  
&\leq & 
B\alpha \;\leq\;  L^{}_{f,\beta}\,\, g_{\max} C_{\beta}(K,K')\, 
 \frac{\log N}{Nh^2} \,\mathbf{1}_{N}\,,  
  \label{constrPrime}\\
 &&D^T\alpha \,\leq\, C_\alpha h\, \mathbf{1}_{m_h}
\,,  \label{constrSumAlpV} \\
  0&\leq&\alpha 
  \,. \label{constr2}
\end{eqnarray}

There is one positive parameter $C_\alpha$ in the constraints
(\ref{constrSumAlp}) and (\ref{constrSumAlpV}):
its value will be discussed in~Section~\ref{secmain}.
Moreover, the following notations have been introduced:
\begin{eqnarray}
X_0&\triangleq& 0, \quad X_{N+1}\triangleq 1, \label{X0X1defs}\\
\mathbf{1}_N &\triangleq& (1, 1,\dots, 1)^T\in \R^N \label{bf1Ndef}\\
A &\triangleq& \left\| K_h(X_i\,,X_j) \right\|_{i,j=1,\dots,N}\label{AmatrDef}\\
B &\triangleq& \left\|  \left. \frac{d}{dx} K_h(x,X_j)\right|_{x=X_i}
\right\|_{i,j=1,\dots,N}
    \label{BmatrDef}\\
D &\triangleq& \left\| \mathbf{1}\{ (j-1)/m_h\leq X_i < j/m_h\} 
			\right\|_{i=1,\dots,N;\, j=1,\dots,m_h} 
    \label{DmatrDef}\\
Y &\triangleq& (Y_1,\dots,Y_N)^T.   \label{YvecDef}
\end{eqnarray}

\section{Preliminary results}
\label{preliresult}

The basic assumptions on the unknown boundary function are:
\begin{itemize}
\item[A1.] $0< f_{\min} \leq f(x) \leq f_{\max} <\infty$, for all $x\in [0,1]$,
 \item[A2.] $|f(x)-f(y)|\leq L^{}_{f,\beta}\, |x-y|^\beta$, for all $x,y \in [0,1]$,\,
with $\, L^{}_{f,\beta}\,<\infty$ and $0<\beta\leq 1$.
\end{itemize}
\medskip

\noindent The following assumptions on the kernel function are
introduced:
\begin{itemize}
  \item[B1.] $K:\R\to[0,\infty)$ has a compact support:
   $\mathop{\mathrm{supp}}_{t\in\R}K(t)=[-1,1]$,
  \item[B2.] $\displaystyle \int_{-1}^1 K(t)\, dt =1 $,
  \item[B3.] $K$ is three times continuously differentiable.   
\end{itemize}
Note, that $g_{\max}=2$  for any unimodal even kernel $K(\cdot)$ meeting conditions B1--B2.
We quote two preliminary results on the estimator $\widehat f_N$.
First, the surface of the related estimated support $\widehat S_N$
is approximatively $\sum_{i=1}^N \alpha_i$. Second, the function
$\widehat f_N$ is Lispchitzian.  
Proofs are postponed to Subsection~\ref{subsec:proofpreli}.

\begin{lemma}\label{LemSurfApprox} 
Suppose B1, B2 are verified and $0<h<1/4$.
Moreover, let conditions (\ref{constrSumAlp}) and (\ref{constr2S}) hold true for $m_h=\lfloor{h^{-1}}\rfloor$. 
Then the surface of the estimated support (\ref{defSesti}) meets the following inequality:
\begin{equation}\label{SurfApproxUB}
-2 C_{\alpha} K_{\max} h \leq
 \int_0^1 \widehat{f}_N(x)\, dx -  \sum_{i=1}^N \alpha_i 
  \leq 4 C_{\alpha} (g_{\max}-1) K_{\max} h \,.
\end{equation}
\end{lemma}

\begin{Remark}\label{rem:HalfUsed}
In fact, only 
one part of Lemma \ref{LemSurfApprox} is used in what follows, that is
the  upper bound for the estimator surface 
given by the right hand side (\ref{SurfApproxUB}).
\end{Remark}
\begin{Remark}\label{rem:Kneg}
Lemma \ref{LemSurfApprox} as well as the further results may be easily extended to basic kernels $K(\cdot)$
having also negative values: then $K_{\max}\triangleq\max |K(t)|$, 
and $g(x)>0\,\,\forall\, x\in[0,1]$ should be additionally assumed.
\end{Remark}
\begin{lemma}\label{RegLemma}
Suppose A1 and B1--B3 are verified.
Let estimator $\widehat{f}_N$ be defined by LP  (\ref{IPgoal})--(\ref{constr2}).
Moreover, let $h\to 0$ as $N\to\infty$ such that
\begin{equation}
\lim_{N\to\infty} \frac{\log N}{N h} =0
\,.
\end{equation}
Then, there exists almost surely finite
$N_4=N_4(\omega)$ such that for any $N\geq N_4$
the Lipschitz constant for the estimator $\widehat{f}_N$ over the interval $[0,1]$ 
is bounded as follows:
\begin{eqnarray}\label{eq:RegLemma}
L_{\widehat{f}_N} &\triangleq&
    \max_{x\in[0,1]} |\widehat{f}_N^{\;\prime}(x)|
\\ \label{eq:RegLemma1}
\label{eq:RegLemmaAdd}
    &\leq &2 L^{}_{f,\beta}\,\, g_{\max} C_{\beta}(K,K') \frac{\log N}{Nh^2}.
\end{eqnarray}
\end{lemma}

\begin{Remark}\label{rem:reserve}
As it can be seen from the proof of Lemma~\ref{RegLemma}, namely from (\ref{eq:ApplAuxLemma1})--(\ref{eq:ApplAuxLemma2}), one might slightly decrease the number
of constraints (\ref{constrPrimeS}) on the estimator derivative (\ref{IPgoalS})--(\ref{constr2S}).
In fact, one could impose those type of constraints not at each point $X_i$, $i=1,\dots,N$:
It would be enough to do at the points with the distance $O\left((h\log N/N)^{1/2}\right)$ between them,
or at least $o\left((h\log N/N)^{1/2}\right)$ in order to keep the same Lipschitz constant for $\widehat{f}_N$
as is given by Lemma~\ref{RegLemma}.
\end{Remark}

\noindent It appears that the estimator $\widehat{f}_N$ being the solution
to the
optimization problem (\ref{IPgoalS})--(\ref{constr2S}) or to its equivalent
LP version (\ref{IPgoal})--(\ref{constr2}) defines the kernel estimator
of the support covering all the points $(X_i, Y_i)$ 
and, approximately, having the smallest surface, up to the term $O(h)$
specified in Lemma \ref{LemSurfApprox}.
Moreover, constraints (\ref{constrPrimeS})--(\ref{constrSumAlp}) or (\ref{constrPrime})--(\ref{constrSumAlpV})
ensure $\widehat{f}_N\in\Sigma_{[0,1]}(1,L_{\widehat{f}_N})$
with a particular Lipschitz constant $L_{\widehat{f}_N}$
given in Lemma~\ref{RegLemma}.
The constraint $\alpha_i\geq 0$ for all $i=1,\dots,N$ ensures that
$\widehat{f}_N(x) \geq 0$ for all $x\in[0,1]$ since the basic kernel $K$ is chosen to be non-negative;
this seems to be natural for function $f(\cdot)$ is positive.  
Finally, note that the above described
estimator (\ref{estimator}),
(\ref{IPgoal})--(\ref{constr2}) may be treated as
the approximation to Maximum Likelihood Estimate related to the estimation
family~(\ref{estimator}); see {\sc Bouchard} {\it et al}~\cite{BGIN,BGINrep} for the demonstration.

\section{Main results}
\label{secmain}

In the following theorem, the consistency and the convergence rate of the 
estimator towards the true frontier
is established with respect to the $L_1$ norm on the
$[0,1]$ interval.

\begin{Theo}\label{Th1}
Let the above mentioned assumptions A1, A2 and B1--B3 hold true and the estimator
parameter $C_{\alpha}>6 f_{\max}$. Moreover, let $h\to 0$ as $N\to\infty$ such that
\begin{equation}
\label{eq:limsup}
\liminf_{N\to\infty} \frac{\log N}{N h^{1+\beta}} > \rho>0
\,,\quad
\lim_{N\to\infty} \frac{\log N}{N h^{1+\beta/2}} =0\,.
\end{equation}
Then estimator (\ref{estimator})--(\ref{constr2}) has the
following asymptotic properties\/:
\begin{equation}\label{L1rate}
\|\widehat{f}_N -f\|_{1} \leq 
\left(
C_{12}(\beta) h^{\beta} + 
2C_4(\beta) h^{-2} (\log N/N)^\frac{2+\beta}{1+\beta}
\right) (1+o(1))
\quad \mathrm{a.s.}
%
%
%
\end{equation}
with
\begin{equation}\label{C12}
C_{12}(\beta) \triangleq 2L^{}_{f,\beta}\,\, g_{\max}\,C_{\beta}(K,K')+ 4 C_{\alpha} (g_{\max}-1) K_{\max} \mathbf{1}\{\beta=1\} 
\end{equation}
and
\begin{equation}\label{ConstC4}
C_4(\beta) \triangleq 2 L^{}_{f,\beta} \left[
\left( \frac{2C_f}{L^{}_{f,\beta}\,}\right)^\frac{\beta}{1+\beta} \left(\frac{1}{\rho} \right)^{\frac{2}{1+\beta}} +
 g_{\max}  C_{\beta}(K,K') \left(\frac{2C_f}{L^{}_{f,\beta}\,}\right)^\frac{1}{1+\beta}\,\right]\,.
\end{equation}
\end{Theo}
\begin{Coro}\label{CorTh1}
The maximum rate of convergence which is guaranteed by Theorem
\ref{Th1}
$$
\|\widehat{f}_N -f\|_{1} = O \left((\log{N}/N)^\frac{\beta}{1+\beta}\right)
\quad \mathrm{a.s.}
$$
is attained for $h$ meeting the following asymptotics:
\begin{equation}\label{opth}
  h \sim  \widetilde{\rho}  \left(\frac{\log{N}}{N}\right)^\frac{1}{1+\beta}, \quad 0<\widetilde{\rho}<\rho^{-\frac{1}{1+\beta}}\,,
\end{equation}
which reduces the upper bound (\ref{L1rate}) to 
\begin{equation}\label{optimL1rate}
\limsup_{N\to\infty}  \left(\frac{\log N}{N}\right)^{-\frac{\beta}{1+\beta}}
\|\widehat{f}_N -f\|_{1} \leq 
C_{12}(\beta) \widetilde{\rho}^{\,\beta} + 
2C_4(\beta) \widetilde{\rho}^{\,-2} 
\quad \mathrm{a.s.}
\end{equation}
\end{Coro}
\medskip
Let us highlight that~(\ref{optimL1rate}) shows that $\widehat f_N$
reaches (up to a logarithmic factor) the minimax $L_1$ rate
for Lipschitz frontier $f$, see
{\sc Korostelev} \& {\sc Tsybakov}~\cite{KorTsy2}, Theorem~4.1.1.
\begin{Remark}\label{rem:extend40}
The second condition in (\ref{eq:limsup}) may be extended to
\begin{equation}\label{eq:extend40}
\lim_{N\to\infty} \frac{\log N}{N h^{1+\beta/2}} <\infty
\end{equation}
which leads to another, more general formula for constants in (\ref{L1rate})--(\ref{ConstC4}).
\end{Remark}
\medskip


\section{Proofs}
\label{proofs}

The proof of Theorem \ref{Th1} which is given in Subsection
\ref{PrTh1} is based on both upper and lower bounds derived in 
Subsection~\ref{subsec:UB} and Subsection~\ref{subsec:LB} respectively.  
When proving these bounds, we assume that the sequence of the sample $X$-points
$(X_i)_{i=1,\dots,N}$ is already increase ordered,
without changing notation from $X_i$ to $X_{(i)}$ for the sake of
simplicity, that is
\begin{equation}\label{Xorder}
X_i\leq X_{i+1}\,,\quad
\forall\, i
\,.
\end{equation}
We essentially apply the uniform asymptotic bound $O(\log{N}/N)$ on
$\Delta X_i \triangleq X_i - X_{i-1}$ proved in
auxiliary Lemma \ref{CXlemma}.
Before that, we prove in Subsection~\ref{subsec:proofpreli} the
two preliminary results.  

\subsection{Proof of preliminary results}
\label{subsec:proofpreli}

\noindent\textbf{Proof of Lemma \ref{LemSurfApprox}.}
Note that definitions (\ref{estimator})--(\ref{Gcorrector}) imply the following decomposition.
\begin{eqnarray}\label{SurfEstima}
\int\limits_0^1 \widehat{f}_N(x)\, dx &=& \left( \int\limits_0^h +  \int\limits_h^{1-h} +  \int\limits_{1-h}^1  \right) \widehat{f}_N(x)\, dx
 \\ \label{SurfEstima1}
 &=&\int\limits_0^1 \sum_{i=1}^N \alpha_i \,\frac{1}{h}\, K\!\left(\frac{x-X_i}{h}\right) dx
 \\ \label{SurfEstima2}
& +& \sum_{i=1}^N \alpha_i  \left( \int_0^h +  \int_{1-h}^1  \right) 
\frac{g(x)-1}{h} \, K\!\left(\frac{x-X_i}{h}\right)  dx
\,.
\end{eqnarray}
Since $\alpha_i$ and kernel $K$ are non-negative, it follows that
\begin{equation}\label{SurfApproxK}
\int\limits_0^1 \sum_{i=1}^N \alpha_i \,\frac{1}{h}\, K\!\left(\frac{x-X_i}{h}\right) dx
\leq   \sum_{i=1}^N \alpha_i \int\limits_\R \frac{1}{h}\, K\!\left(\frac{x-X_i}{h}\right) dx
= \sum_{i=1}^N \alpha_i \,
\end{equation}
and therefore,
\begin{eqnarray}\label{SurfApproxUBpr}
 \int\limits_0^1 \widehat{f}_N(x)\, dx - \sum_{i=1}^N \alpha_i 
&\leq& \frac{g_{\max}-1}{h}  K_{\max}  \sum_{i=1}^N \alpha_i  \left( \int\limits_0^h +  \int\limits_{1-h}^1  \right) 
\! \mathbf{1}\{|x-X_i|\leq{h}\}  dx
 \\ \label{SurfApproxUBpr1}
 &\leq& (g_{\max}-1) K_{\max} \left(
  \sum_{i=1}^N \alpha_i \, \mathbf{1}\{0\leq X_i\leq 2h\} \right. 
  \\ \label{SurfApproxUBpr2}
 && \left. \phantom{(g_{\max}-1) K_{\max} \left( \right. }
   + \, \sum_{i=1}^N \alpha_i \, \mathbf{1}\{1-2h\leq X_i\leq 1\} \right) 
  \\ \label{SurfApproxUBpr3}
  &\leq&  (g_{\max}-1) K_{\max} 4 C_{\alpha} h \,.
\end{eqnarray}
The inequality (\ref{SurfApproxUBpr3}) follows from (\ref{constrSumAlp}) since
both intervals
in (\ref{SurfApproxUBpr1})--(\ref{SurfApproxUBpr2}) are of the length $2h$
and thus may be covered by two related intervals of the form $[(j-1)/m_h, j/m_h]$ in (\ref{constrSumAlp}).
Consequently, we have proved the upper bound for the difference in the left hand side~(\ref{SurfApproxUBpr}).
The lower bound is proved in the same manner. Indeed, decomposition (\ref{SurfEstima})--(\ref{SurfEstima2})
implies, since the term (\ref{SurfEstima2}) is non-negative,
\begin{eqnarray}\label{SurfApproxLBpr}
 \int\limits_0^1 \widehat{f}_N(x)\, dx - \sum_{i=1}^N \alpha_i 
&\geq& - \sum_{i=1}^N \alpha_i  \left( \,\int\limits_{-h}^0 +  \int\limits_1^{1+h} \, \right)  \frac{1}{h} 
\, K\!\left(\frac{x-X_i}{h} \right) dx
 \\ \label{SurfApproxLBpr1}
 &\geq& - K_{\max} \left(
 \sum_{i=1}^N \alpha_i \, \mathbf{1}\{0\leq X_i\leq h\} \right. 
  \\ \label{SurfApproxLBpr2}
 && \left. \phantom{ K_{\max} \left(aaa \right. }
   + \, \sum_{i=1}^N \alpha_i \, \mathbf{1}\{1-h\leq X_i\leq 1\} \right) 
  \\ \label{SurfApproxLBpr3}
  &\geq&  - K_{\max}\, 2 C_{\alpha} h \,.
\end{eqnarray}
This completes the proof of Lemma~\ref{LemSurfApprox}.
\CQFD

\noindent\textbf{Proof of Lemma \ref{RegLemma}.}
Remind that we assume (\ref{Xorder}). 
By applying auxiliary Lemma~\ref{CXlemma} and Lemma~\ref{AuxLemma}
we first arrive at
\begin{eqnarray}\label{eq:ApplAuxLemma}
&&    \max_{x\in[0,1]} |\widehat{f}_N^{\;\prime}(x)|\\
&=& \max_{1\leq i\leq N+1}\, \max_{x\in[X_{i-1},X_i]} |\widehat{f}_N^{\;\prime}(x)|
\\ \label{eq:ApplAuxLemma1}
    &\leq&  L^{}_{f,\beta}\,\, g_{\max} C_{\beta}(K,K')\frac{\log N}{Nh^2}
     +\frac{1}{8} \max_{1\leq i\leq N+1} \left[(X_i-X_{i-1})^2
             \max_{x\in[X_{i-1},X_i]} |\widehat{f}_N^{\;\prime\prime\prime}(x)|
                                    \right]
\\ \label{eq:ApplAuxLemma2}
    &\leq&  L^{}_{f,\beta}\,\, g_{\max} C_{\beta}(K,K')\frac{\log N}{Nh^2}
     +\frac{1}{8} \left(C_X \frac{\log{N}}{N}\right)^2
            \max_{x\in[0,1]} |\widehat{f}_N^{\;\prime\prime\prime}(x)|\,,
\end{eqnarray}
with $C_X>4 C_f/f_{\min}$.  
The maximum term in (\ref{eq:ApplAuxLemma2}) is bounded as follows: for any
$x\in[0,1]$
\begin{eqnarray}\label{eq:MaxTermBound}
|\widehat{f}_N^{\;\prime\prime\prime}(x)|  
&\leq& \sum_{i=1}^N \alpha_i 
\left| \frac{d^3}{dx^3}\,K_h( x,X_i)
\right|
\\ \label{eq:MaxTermBound1}
    &\leq& \sup_{u,v} \left|\frac{\partial^3}{\partial v^3}\,K_h(v,u) \right|
        \cdot     \sum_{i=1}^N \alpha_i \,\mathbf{1}\{|x-X_i|\leq h\}
\\ \label{eq:MaxTermBound2}
    &\leq& g_{\max} L_{\widetilde{K}^{\prime\prime}} h^{-4} \cdot 3C_{\alpha} h \,,
\end{eqnarray}
since (see Lemma~\ref{LipConstsLem} for the detailed demonstration) 
\begin{equation}\label{eq:LK2primHb}
 \sup_{u,v} \left|\frac{\partial^3}{\partial v^3}\,K_h(v,u) \right|
 \leq g_{\max} L_{\widetilde{K}^{\prime\prime}} h^{-4}
\end{equation}
where
\begin{equation}\label{eq:LK2primHa}
L_{\widetilde{K}^{\prime\prime}} \triangleq 
 L_{K''} + 3L_{K'} K_{\max}g_{\max} 
 + L_Kg_{\max}\left(3L_K+10K_{\max}^2g_{\max}\right) +6K_{\max}^4g_{\max}^3
   \,.
\end{equation}
Substituting (\ref{eq:MaxTermBound}), (\ref{eq:MaxTermBound2}) into 
(\ref{eq:ApplAuxLemma2}) yields
\begin{eqnarray}
    \max_{x\in[0,1]} |\widehat{f}_N^{\;\prime}(x)|
    &\leq&
      L^{}_{f,\beta}\, g_{\max} C_{\beta}(K,K') \frac{\log N}{Nh^2}
     +  \frac{3}{8}\,g_{\max}  L_{\widetilde{K}^{\prime\prime}}  C_{\alpha} \left(C_X \frac{\log{N}}{N h^2}\right)^2   h\\
& \leq &2 L^{}_{f,\beta}\,\, g_{\max} C_{\beta}(K,K') \frac{\log N}{Nh^2}
\end{eqnarray}
under additional assumption (which hold true for all sufficiently large $N$):
\begin{equation}\label{eq:hAddAssump}
    h\geq    \frac{ 3C_X^{\,2}C_{\alpha}\, L_{\widetilde{K}^{\prime\prime}}\, \log{N}}
                { 8 L^{}_{f,\beta}\,C_{\beta}(K,K')\, N} \,.
\end{equation}
The result follows.  
\CQFD\medskip
\subsection{Upper bound for $\widehat{f}_N$ in terms of $J^*_P$}
\label{subsec:UB}

\begin{lemma}\label{Lm1Th1}
Let the assumptions of Theorem \ref{Th1}
hold true. Then for any finite 
\begin{equation}\label{GammaCondi}
\gamma >L^{}_{f,\beta}\, g_{\max}C_{\beta}(K)
\end{equation} 
and
almost all $\omega\in\Omega$ there exist finite
numbers $N_1=N_1(\omega,\gamma)$ such that
for all $N\geq N_1$ the LP (\ref{IPgoal})--(\ref{constr2}) is solvable and
\begin{equation}\label{UB}
  J^*_P \leq C_f +\gamma h^\beta\,.
\end{equation}
\end{lemma}
\medskip

\noindent\textbf{Proof of Lemma \ref{Lm1Th1}.}
Consider arbitrary $N\geq N_0(\omega)$ with $N_0(\omega)$ from Lemma \ref{CXlemma}.
Introduce function $f_\gamma(u)=f(u)+\gamma h^\beta$
and pseudo-estimators
\begin{equation}\label{pseudoestimates}
\widetilde{\alpha}_i
 = \frac{1+\delta_{i1}}{2} \int_{X_{i-1}}^{X_{i}} f_\gamma(u)\,du
 + \frac{1+\delta_{iN}}{2} \int_{X_{i}}^{X_{i+1}} f_\gamma(u)\,du
 \,,\quad i=1,\dots,N
\end{equation}
where $\delta_{ij}$ stands for Kronecker symbol. Below we
demonstrate that condition (\ref{GammaCondi})
 ensures the vector of pseudo-estimators
$\widetilde{\alpha}=(\widetilde{\alpha}_1\,\dots,\widetilde{\alpha}_N)^T$
to be an admissible point for the LP
(\ref{IPgoal})--(\ref{constr2}),
for any sufficiently large $N$. This implies solvability of the LP (\ref{IPgoal})--
(\ref{constr2}) and
\begin{equation}\label{JlpLessLtilde}
J_P^* \leq \sum_{i=1}^N \widetilde{\alpha}_i
= \int_0^1 (f(u)+\gamma h^\beta)\,du
= C_f +\gamma h^\beta\,.
\end{equation}
Let $C_X>4 C_f/f_{\min}$.  
For the sake of simplicity, we impose the
additional assumptions 
\begin{equation}\label{eq:AddAssumptions}
h^\beta \leq \frac{\log{N}}{\rho Nh} 
    \leq  \min\left\{\frac{f_{\max}}{\gamma},\frac{1}{\rho C_X}\right\},
\end{equation}
which hold true for all $N$ large enough.  

\noindent
1. First, we prove constraints (\ref{constr1S}) under
$\alpha_i = \widetilde{\alpha}_i$, $i=1,\dots,N$. For arbitrary $x\in[0,1]$,
\begin{eqnarray}\label{ProveConstraints1}
\widetilde{f}_N(x) &\triangleq&
     \displaystyle \sum_{i=1}^N \widetilde{\alpha}_i K_h(x,X_i)
 \\ \label{ProveConstraints11}
&=& \displaystyle \sum_{i=1}^{N+1} \int_{X_{i-1}}^{X_i}
        f_\gamma(u)\,du\,\frac{K_h(x,X_i)+K_h(x,X_{i-1})}{2}
 \\ \label{ProveConstraints12}
    && +\, \frac{1}{2}\int_0^{X_1} f_\gamma(u)\,du \left(K_h(x,X_1)-K_h(x,0) \right)
 \\ \label{ProveConstraints13}
    && +\, \frac{1}{2}\int_{X_N}^1 f_\gamma(u)\,du \left(K_h(x,X_N)-K_h(x,1) \right)
 \\ \label{ProveConstraints14}
&=& \int_{0}^{1} f_\gamma(u)\, K_h(x,u)\,du
 \\ \label{ProveConstraints15}
    && +\displaystyle \sum_{i=1}^{N+1} \int\limits_{X_{i-1}}^{X_i}\!\!
        f_\gamma(u)\left(\!\frac{K_h(x,X_i)\!+\!K_h(x,X_{i-1})}{2} -K_h(x,u)\!\!
                    \right)\!du
 \\ \label{ProveConstraints16}
    && +\, \frac{1}{2}\int_0^{X_1} f_\gamma(u)\,du \left(K_h(x,X_1)-K_h(x,0) \right)
 \\ \label{ProveConstraints17}
    && +\, \frac{1}{2}\int_{X_N}^1 f_\gamma(u)\,du \left(K_h(x,X_N)-K_h(x,1) \right)\,.
\end{eqnarray}
Now we separately bound each of the summands
 (\ref{ProveConstraints14})--(\ref{ProveConstraints17}) from below.
 Due to  (\ref{IntCorKern1}), the main term (\ref{ProveConstraints14})
 is bounded as follows:
\begin{eqnarray}\label{ProveConstraints14sep}
 \int_{0}^{1} f_\gamma(u)\, K_h(x,u)\,du
  &=& f(x) + \gamma h^\beta + \int_{0}^{1} (f(u)-f(x))\, K_h(x,u)\,du
 \\ \label{ProveConstraints14sep1}
&\geq& f(x) + (\gamma - L^{}_{f,\beta}\,\, g_{\max}C_{\beta}(K)) h^\beta\,.
\end{eqnarray}
The $i$-th summand from (\ref{ProveConstraints15}) is decomposed and then bounded
basing on trapezium formula error as follows:
\begin{eqnarray}\label{ProveConstraints15sep}
&&\int_{X_{i-1}}^{X_i}
        f_\gamma(u)\left(\frac{K_h(x,X_i)+K_h(x,X_{i-1})}{2} -K_h(x,u)
                    \right)du
 \\ \label{ProveConstraints15sep1}
&\geq& f_\gamma(x)\int_{X_{i-1}}^{X_i}\left(\frac{K_h(x,X_i)+K_h(x,X_{i-1})}{2} -K_h(x,u)
                    \right)du
 \\ \label{ProveConstraints15sep2}
 && -\int\limits_{X_{i-1}}^{X_i}\!\! |f_\gamma(u)-f_\gamma(x)| 
        \left|\frac{K_h(x,X_i)+K_h(x,X_{i-1})}{2} -K_h(x,u)
                    \right|du
 \\ \label{ProveConstraints15sep3}
&\geq& -(f_{\max}+\gamma h^\beta) \, \frac{(X_i-X_{i-1})^3}{12} \max_{u\in[0,1]}
 \left|\frac{\partial^2 K_h(x,u)}{\partial u^2}\right| \,\mathbf{1}\{|x-X_i|\leq 2h\}
 \\ \label{ProveConstraints15sep4}
 && - L^{}_{f,\beta}\, \int_{X_{i-1}}^{X_i} |u-x|^\beta\,\mathbf{1}\{|x-X_i|\leq 2h\}\frac{g_{\max}L_K}{2h^2}
 \,[(u-X_{i-1})+(X_i-u)]\,du.  
\label{ProveConstraints15sep5}
\end{eqnarray}
By applying Lemma~\ref{CXlemma}, the first term is bounded as follows
\begin{equation}
\label{firstterm}
(\ref{ProveConstraints15sep3})
\geq
-\left( C_X \frac{\log{N}}{N}\right)^2
 \frac{f_{\max}\,g_{\max}L_{K^{\,\prime}}}{6h^3} \,(X_i-X_{i-1})\,\mathbf{1}\{|x-X_i|\leq 2h\},
\end{equation}
and the second one is bounded by:
\begin{equation}\label{secondterm}
(\ref{ProveConstraints15sep4})
\geq
   -\frac{g_{\max} L^{}_{f,\beta}\, L_K}{2h^2} C_X\frac{\log N}{N}\mathbf{1}\{|x-X_i|\leq 2h\}
\int_{X_{i-1}}^{X_i} |u-x|^\beta du\,.
\end{equation}
Moreover, from Lemma~\ref{CXlemma}, one can show first that
$$
\sum_{i=1}^{N+1} \mathbf{1}\{|x-X_i|\leq 2h\} (X_i -X_{i-1}) 
\leq 4h + \frac{C_X \log N}{N},
$$
and second that
\begin{eqnarray}
&&\sum_{i=1}^{N+1} \mathbf{1}\{|x-X_i|\leq 2h\} \int_{X_{i-1}}^{X_i} |u-x|^\beta du\\
&\leq& \int_{x-2h-C_X(\log N)/N}^{x+2h}|u-x|^\beta du\\
&\leq& \left(4h+C_X\frac{\log N}{N}\right) \max_{v\in[-2h-C_X(\log N)/N, 2h]}|v|^\beta\\
&\leq& \left(4h+C_X\frac{\log N}{N}\right) \left(2h+C_X\frac{\log N}{N}\right)^\beta.
\end{eqnarray}
Thus, we arrive at the bound for the sum (\ref{ProveConstraints15}) as follows:
\begin{eqnarray}\label{ProveConstraints15sum}
&&\displaystyle \sum_{i=1}^{N+1} \int_{X_{i-1}}^{X_i}
        f_\gamma(u)\left(\frac{K_h(x,X_i)+K_h(x,X_{i-1})}{2} -K_h(x,u)
                    \right)du
\\ \label{ProveConstraints15sum51}
&\geq&
- g_{\max} C_X \frac{\log N}{N h}\left(4 + C_X\frac{\log N}{Nh}\right)
\left(\frac{C_Xf_{\max}L_{K'}}{6} \frac{\log N}{N h}
 \right.
 \\  \label{ProveConstraints15sum512}
 && \phantom{- g_{\max} C_X \frac{\log N}{N h}\left(4 + C_X\frac{\log N}{Nh}\right)\left(\right)}
  \left.
+\, \frac{L^{}_{f,\beta}\, L_K}{2}h^\beta\left(2+C_X \frac{\log N}{Nh}\right)^\beta\right)
\\ \label{ProveConstraints15sum52}
&\geq&
- \frac{5}{6} g_{\max} C_X \left(\frac{\log N}{N h}\right)^2
\left(C_X f_{\max} L_{K'} 
+ 3^{\beta+1} \rho^{-1} L^{}_{f,\beta}\, L_K \right).
\end{eqnarray}
At last, it is similarly demonstrated that
both summands (\ref{ProveConstraints16}) and (\ref{ProveConstraints17})
are bounded above by $O((\log{N}/(N h))^2)$.
For instance, for
 (\ref{ProveConstraints16}), one obtains
\begin{eqnarray}\nonumber
\left|\int_0^{X_1} f_\gamma(u)\,du \left(K_h(x,X_1)-K_h(x,0) \right)\right|
&\leq& (f_{\max}+\gamma h^\beta) X_1 \left|K_h(x,X_1)-K_h(x,0)\right|\\
&\leq& 2f_{\max}\, g_{\max} L_K \left( C_X \frac{\log{N}}{Nh}\right)^2.
\label{ProveConstraints16sum}
\end{eqnarray}
Thus, from (\ref{ProveConstraints1})--(\ref{ProveConstraints16sum})
it follows for each $j=1,\dots,N$ that
\begin{equation}\label{LastTerms}
\widetilde{f}_N(X_j) \geq f(X_j) + (\gamma - L^{}_{f,\beta}\, \, g_{\max}C_{\beta}(K)) h^\beta 
+O\left(\left(\frac{\log N}{Nh}\right)^2\right)
 \geq Y_j
\end{equation}
for sufficiently large $N\geq N_0(\omega)$ when both inequalities (\ref{eq:AddAssumptions})
and the following one hold true:
\begin{equation}\label{eq:LastCond}
 \gamma - L^{}_{f,\beta}\,\, g_{\max}C_{\beta}(K)\geq
\frac{5}{6}\, g_{\max} C_X \left(\frac{\log N}{N h^{1+\beta/2}}\right)^2
\left(C_X f_{\max}\left(L_{K'}+\frac{12L_K}{5}\right)
+ 3^{\beta+1}  \frac{L^{}_{f,\beta}\, L_K}{\rho} \right).
\end{equation}
\medskip

\noindent
2. Similarly, constraints (\ref{constrPrimeS}) hold true under
$\alpha_i = \widetilde{\alpha}_i$, $i=1,\dots,N$. Indeed, for arbitrary $x\in[0,1]$,
we now have to bound  the absolute value of
\begin{equation}\label{ProveConstrPrimeS}
\widetilde{f}_N^{\;\prime}(x) =
     \displaystyle \sum_{i=1}^N \widetilde{\alpha}_i \,\frac{d}{dx}\,K_h(x,X_i) =
     \displaystyle \sum_{i=1}^N \widetilde{\alpha}_i \,\widetilde{K}_h(x,X_i)
\end{equation}
instead of (\ref{ProveConstraints1}). 
Here
\begin{equation}\label{DerivKerC}
\widetilde{K}_h(x,u) \triangleq \frac{\partial}{\partial x}\,K_h(x,u)
 \end{equation}
(see Subsection~\ref{subsecCorKernel}) with the following upper bound 
\begin{equation}\label{TildeKerUB}
\left| \widetilde{K}_h(x,u) \right|
\leq h^{-2} g_{\max} \left\{ \left|K'\left(\frac{x-u}{h}\right)\right| + g_{\max} \,K_{\max} \left|K\left(\frac{x-u}{h}\right) \right| \right\}\,.
\end{equation}
deduced form (\ref{CorKernLeftG0}), (\ref{CorKernLeftG1}).
Hence,
one may repeat the arguments of (\ref{ProveConstraints11})--(\ref{ProveConstraints17})
by changing $K_h$ for $\widetilde{K}_h$. Therefore,
all the rates from (\ref{ProveConstraints15sep})--(\ref{LastTerms})
should be divided by $h$, while the absolute value of the
main term of decomposition, due to (\ref{IntCorKernDerx1}),
 is bounded as follows:
\begin{eqnarray}\label{ProveConstrPrimeSmain} 
 \left|\int_{0}^{1} f_\gamma(u)\,\widetilde{K}_h(x,u) \,du \right|
  &=& \left|\int_{0}^{1} (f(u)-f(x))\,\frac{\partial}{\partial x}\,K_h(x,u)\,du \right|
 \\ \label{ProveConstrPrimeSmain1}
&\leq& L^{}_{f,\beta}\,\, g_{\max} C_{\beta}(K,K')h^{\beta-1}\,,
\end{eqnarray}
instead of (\ref{ProveConstraints14sep})--(\ref{ProveConstraints14sep1}).
Remind the definition (\ref{Cbeta})--(\ref{Cbeta2}) for $C_{\beta}(K,K')$ which follows from (\ref{TildeKerUB}).
Thus, for sufficiently large $N\geq N_0(\omega)$
and for each $X_j$ we arrive at
\begin{equation}\label{ProveConstrPrimeS1}
\left|\widetilde{f}_N^{\;\prime}(X_j)\right| \leq L^{}_{f,\beta}\,\, g_{\max} C_{\beta}(K,K')
h^{\beta-1}+ O\left(\frac{\log^2 N}{N^2h^{3}}\right) 
\leq  L^{}_{f,\beta}\,\, g_{\max} C_{\beta}(K,K')\frac{\log N}{\rho Nh^2}.
\end{equation}
Namely, inequality (\ref{ProveConstrPrimeS1}) holds true almost surely for all those
$N\geq N_0(\omega)$ such that~(\ref{eq:AddAssumptions}) is verified and
\begin{eqnarray}\label{eq:AddAssumptions2}
L^{}_{f,\beta}\, C_{\beta}(K,K')\left(\frac{\log N}{h^{\beta+1}N}-1\right)
&\geq&
\frac{5}{6}\, g_{\max} C_X \left(\frac{\log N}{N h^{1+\beta/2}}\right)^2
\\ \label{eq:AddAssumptions21}
&& \cdot
\left(C_X f_{\max}\left(L_{\widetilde{K}'}+\frac{12L_{\widetilde{K}}}{5}\right)
%
+\, 3^{\beta+1} \frac{L^{}_{f,\beta}\, L_{\widetilde{K}}}{\rho} 
\phantom{\frac{1}{2}} \!\!\!\!\! \right)
\end{eqnarray} 
where  (see Lemma~\ref{LipConstsLem} for the detailed demonstration)
\begin{equation}\label{eq:AddAssumptions22}
L_{\widetilde{K}} \triangleq L_{K'} +L_K\,g_{\max}K_{\max}
\,,\quad L_{\widetilde{K}'} \triangleq L_{K''} +L_{K'}\,g_{\max}K_{\max} \,.
\end{equation} 
\medskip

\noindent
3. Finally, the constraints (\ref{constrSumAlp}) with
\begin{equation}\label{CalphaValue}
C_\alpha\geq 6 f_{\max}
\end{equation}
also hold true under $\alpha_i = \widetilde{\alpha}_i$, $i=1,\dots,N$.
Indeed, by Lemma \ref{CXlemma} the following inequalities hold \textrm{a.s.} for all $N\geq N_0(\omega)$ 
and for each $j=1,\dots,m_h$, where $m_h=\lfloor{h^{-1}}\rfloor$ :
\begin{eqnarray}\label{eq:FinalConstrCheck}
\sum_{i=1}^N \widetilde{\alpha}_i \, \mathbf{1}\{ (j-1)/m_h\leq X_i < j/m_h\} 
&\leq& (f_{\max}+\gamma h^\beta) \left( 1/m_h + 2C_X \frac{\log{N}}{N}\right)
\\
    &\leq& 6 f_{\max} h,
\end{eqnarray}
under additional assumptions (\ref{eq:AddAssumptions}).
Thus, constraints (\ref{constrSumAlp}) are fulfilled under (\ref{CalphaValue}) almost sure,
for any sufficiently large $N$.
\medskip

\noindent
4. Since all $\widetilde{\alpha}_i\geq 0$, constraints (\ref{constr2S}) hold true, 
and  Lemma~\ref{Lm1Th1} is proved.
\CQFD
\medskip

\begin{Remark}\label{rem:lognu}
By applying Lemma~\ref{SumDX3lemma}, under additional assumptions (\ref{eq:hNN-1}) on $h$, 
one may ameliorate the related bounds in (\ref{ProveConstraints15sum})--(\ref{eq:LastCond}) 
and (\ref{eq:AddAssumptions2})--(\ref{eq:AddAssumptions21}).
Indeed, Lemma~\ref{SumDX3lemma}, being applied with its parameter $\nu\in(1,2)$, states that
\begin{equation}\label{osecondterm}
\sum_{i=1}^{N+1} \mathbf{1}\{|x-X_i|\leq 2h\} (X_i -X_{i-1})^3 = o\left( \frac{h\log^{\nu}N}{N^2}\right)
\end{equation}
hence, the sum of the term (\ref{ProveConstraints15sep3}) is negligible with respect to
that of (\ref{ProveConstraints15sep5}). It means, roughly speaking, that we may remove the term $L_{K'}$
from (\ref{ProveConstraints15sum52}), (\ref{eq:LastCond}) 
as well as $L_{\widetilde{K}'}$ from (\ref{eq:AddAssumptions21}).
However, it does not change much in the main result of the Theorem.
That is why we restrict ourselves to the pointing out this possibility here.
\end{Remark}

\subsection{Lower bound for $\widehat{f}_N$}
\label{subsec:LB}

\begin{lemma}\label{Lm2Th1}
Under the assumptions of Theorem \ref{Th1}, for almost all
$\omega\in\Omega$ there exist finite numbers $N_2(\omega)$ 
such that for any $x\in [0,1]$ and 
for all $N\geq N_2(\omega)$
\begin{equation}\label{LB}
  \widehat{f}_N(x) \geq f(x)- \frac{C_4(\beta)}{h^2} \left(\frac{\log N}{N}\right)^\frac{2+\beta}{1+\beta}
  %
%
\end{equation}
with  constant $C_4(\beta)$ defined in (\ref{ConstC4}).
\end{lemma}
\medskip

%
\noindent\textbf{Proof of Lemma \ref{Lm2Th1}.}
Let us take use of Lemma~\ref{AuxPartiLemma} and its Corollary~\ref{CorAuxPartiLemma}
introducing
\begin{equation}\label{dely}
\delta_y = L^{}_{f,\beta}\,\delta_x^{\,\beta},\quad
\delta_x \triangleq\left(\frac{2C_f\log N}{f_{\min}L^{}_{f,\beta}\,N}\right)^{\frac{1}{1+\beta}}\,.
\end{equation}
Thus, for any $N\geq N_6(\omega)$ and any $x\in[0,1]$ there exists
(with probability one) an integer
$i_k\in\{1,\dots,N\}$ such that
\begin{equation}\label{epsuppose}
 |x-X_{i_k}| \leq \delta_x
 \end{equation}
and
\begin{equation}\label{deltay}
  Y_{i_k}\geq f(X_{i_k}) -\delta_y \,.
\end{equation}
Now, the estimation error at a point $x$ can be expanded as
\begin{eqnarray}\label{ferr1}
  f(x)-\widehat{f}_N(x) &=& \left[f(x)-f(X_{i_k})\right]\\
 \label{ferr2}      &+& \left[f(X_{i_k})- \widehat{f}_N(X_{i_k})\right]\\
 \label{ferr3}      &+& \left[\widehat{f}_N(X_{i_k}) - \widehat{f}_N(x)\right].
\end{eqnarray}
The term in the right hand side (\ref{ferr1}) may be bounded as
follows
\begin{equation}\label{err1}
  \left|f(x)-f(X_{i_k})\right| \leq L^{}_{f,\beta}\,  \left|x-X_{i_k} \right|^{\beta} \leq L^{}_{f,\beta}\, \delta_x^\beta,
\end{equation}
as well as the term (\ref{ferr3})
\begin{equation}\label{err3}
  \left|\widehat{f}_N(X_{i_k}) - \widehat{f}_N(x)\right|
  \leq L_{\widehat{f}_N}  \left|x-X_{i_k} \right| \leq L_{\widehat{f}_N}\, \delta_x,
\end{equation}
with a Lipschitz constant $L_{\widehat{f}_N}\;$ for the function
estimator $\widehat{f}_N(x)$.
Remind that $\widehat{f}_N(X_{i_k})\geq Y_{i_k}$ due to
(\ref{constr1S}) or (\ref{constr1}). Thus, (\ref{deltay}) implies
\begin{equation}\label{err2}
f(X_{i_k})- \widehat{f}_N(X_{i_k}) \leq (Y_{i_k}+\delta_y) -
Y_{i_k} =\delta_y\,.
\end{equation}
Combining all these bounds we obtain from (\ref{ferr1}) that for
all $N\geq N_6(\omega)$,
\begin{equation}\label{ferr12}
f(x)-\widehat{f}_N(x) \leq \delta_y + L^{}_{f,\beta}\,\delta_x^\beta +
L_{\widehat{f}_N} \delta_x\,.
\end{equation}
Therefore, applying Lemma \ref{RegLemma}
 and substituting
 expressions (\ref{dely}) for $\delta_x$ and $\delta_y$ into (\ref{ferr12})
  lead to the lower bound
\begin{eqnarray}\label{LBestim}
\widehat{f}_N(x) &\geq& f(x)- \left( 2L^{}_{f,\beta}\, \delta_x^\beta +
L_{\widehat{f}_N} \delta_x \right)
\\ \label{LBestim1}
&\geq &
f(x)- \frac{C_4(\beta)}{h^2} \left(\frac{\log N}{N}\right)^\frac{2+\beta}{1+\beta}
\end{eqnarray}
for any sufficiently large $N$ (starting from random a.s. finite integer, which does not depend on $x$).
The first inequality in (\ref{eq:limsup}) has been applied here in order to simplify the lower bound.
Lemma \ref{Lm2Th1} is proved.
\CQFD

\subsection{Proof of Theorem \ref{Th1}}\label{PrTh1}
 \medskip \noindent  1. Since $|u|=u-2u\mathbf{1}\{u<0\}$,
 the $L_1$-norm of estimation error can be expanded as
\begin{eqnarray}\label{L1norm1}
  \|\widehat{f}_N -f\|_{1} &=& \int_0^1 \left[\widehat{f}_N(x) -f(x)\right]\,dx
\\ \label{L1norm2}
& +&2\int_0^1 \left[f(x)-\widehat{f}_N(x) \right] \mathbf{1}\! \left\{\widehat{f}_N(x)<f(x)\right\}\,dx.
\end{eqnarray}
 \medskip

 \noindent  2.  Applying Lemmas \ref{LemSurfApprox} and \ref{Lm1Th1} to the right hand side (\ref{L1norm1}) yields
\begin{equation}\label{useUB}
  \limsup_{N\to\infty}\, h^{-\beta}
    \left(\int_0^1 \left[\widehat{f}_N(x) -f(x)\right]\,dx\right)
    \leq \gamma + 4 C_{\alpha} (g_{\max}-1) K_{\max} \mathbf{1}\{\beta=1\}
\quad\mathrm{a.s.}
\end{equation}
Note, that one may fix $\gamma=2L^{}_{f,\beta}\,\, g_{\max}\,C_{\beta}(K)$, for instance.
\medskip

\noindent  3. In order to obtain a similar result for the term (\ref{L1norm2}), note that Lemma \ref{Lm2Th1}
implies
$$
\zeta_N(x,\omega) \triangleq \varepsilon_{LB}^{-1}(N) \left[f(x)-\widehat{f}_N(x) \right]
\leq C_4(\beta)<\infty \quad \mathrm{a.s.}
$$
uniformly with respect to both $x\in[0,1]$ and $N\geq N_2(\omega)$,
with
\begin{equation}\label{epsLB}
  \varepsilon_{LB}(N) \triangleq \frac{1}{h^2}\left(\frac{\log N}{N}\right)^\frac{2+\beta}{1+\beta}.
\end{equation}
Hence, one may apply Fatou lemma, taking into account that
$u\mathbf{1}\{u>0\}$ is a continuous, monotone function:
\begin{eqnarray}
  &&  \limsup_{N\to\infty}\, \varepsilon_{LB}^{-1}(N)
  \int_0^1 \left[f(x)-\widehat{f}_N(x) \right] \mathbf{1}\! \left\{\widehat{f}_N(x)<f(x)\right\}\,dx
   \\
 &\leq&
 \int_0^1 \limsup_{N\to\infty}\, \zeta_N(x,\omega)\, \mathbf{1}\!
    \left\{\zeta_N(x,\omega)>0\right\}\,dx
   \\
 &\leq& C_4(\beta) <\infty \quad \mathrm{a.s.}
\end{eqnarray}

\noindent 4. Thus, the obtained relations together with (\ref{L1norm1}) and
(\ref{L1norm2}) imply (\ref{L1rate}).
Theorem~\ref{Th1} is proved. \CQFD

\section{Appendix}

In Subsection \ref{subsecCorKernel} we establish some properties related to
the corrected kernel. Subsection~\ref{subsecAuxLemsI} presents some
auxiliary lemmas which have been used to prove Theorem~\ref{Th1}.
Finally, we collect in Subsection~\ref{subsecAuxLemsII} some
lemmas dedicated to the proof of Remark~\ref{rem:lognu}.

\subsection{Corrected kernel}
\label{subsecCorKernel}

Let the basic kernel function $K$ be defined as in Section \ref{secmain},
and the bandwidth $h\in(0,1/2)$.
Remind the estimator $\widehat{f}_N$ defined in (\ref{estimator})
as follows:
\begin{equation}
\label{Cestimator}
\left\{
\begin{array}{l}
\widehat{f}_N(x) = \displaystyle \sum_{i=1}^N \alpha_i K_h(x,X_i)
\\
\alpha_i \geq 0,\qquad i=1,\dots,N,
\end{array}
\right.
\end{equation}
where the kernel function
\begin{equation}
\label{CorKernOld}
K_h(x,t)=h^{-1}K((x-t)/h)\qquad \forall\, x\in(h,1-h)
\end{equation}
while
\begin{equation}
\label{CorKernLeft}
K_h(x,t)=h^{-1}K((x-t)/h)\left(\int_{-1}^{x/h}K(t)\,dt \right)^{-1}
\qquad \forall\, x\in[0,h]
\end{equation}
and
\begin{equation}
\label{CorKernRight}
K_h(x,t)=h^{-1}K((x-t)/h)\left(\int_{(x-1)/h}^{1}K(t)\,dt \right)^{-1}
\qquad \forall\, x\in[1-h,1]\,.
\end{equation}
Thus,  the kernel function $K_h(x,t)$ is defined for any $(x,t)\in[0,1]\times\mathbb{R}$, and the estimator (\ref{Cestimator}) is defined for any $x\in[0,1]$
via the kernel $K_h(x,t)$ corrected at the ``boundaries".
One may easily observe that
\begin{equation}\label{IntCorKern}
\int_0^1 K_h(x,u)\,du =1\qquad \forall\, x\in[0,1]
\end{equation}
and, consequently, due to exchanging the integral and the derivative,
\begin{equation}\label{IntCorKernDerx}
\int_0^1 \frac{\partial}{\partial{x}} K_h(x,u)\;du =0\qquad \forall\, x\in[0,1]
\end{equation}
Note, that equation (\ref{IntCorKernDerx}) may also be verified directly.
For instance, on the left boundary, i.e. for $x\in[0,h]$, we have
\begin{equation}\label{CorKernLeftG}
K_h(x,t)=h^{-1}K((x-t)/h)g(x)\,,\qquad g(x)=\left(\int_{-1}^{x/h}K(t)\,dt \right)^{-1}.
\end{equation}
Denoting
\begin{equation}
\label{defKtilde}
\widetilde{K}_h(x,u)=\frac{\partial}{\partial{x}} K_h(x,u)\,,
\end{equation}
we thus have
\begin{equation}
\label{gDer}
g'(x)=-\left(\int_{-1}^{x/h}K(t)\,dt \right)^{-2} h^{-1} K(x/h) = -g^2(x) h^{-1} K(x/h)
\end{equation}
and
\begin{eqnarray}
\label{CorKernLeftG00}
\widetilde{K}_h(x,u)&=& h^{-1}K((x-u)/h)g'(x) + g(x)h^{-2}K'((x-u)/h)\\
&=&  g(x)h^{-1} \left( h^{-1} K'\left(\frac{x-u}{h}\right) 
 -K\left(\frac{x-u}{h}\right) K_h(x,0) \right).
\label{CorKernLeftG10}
\end{eqnarray}
Hence, the integral
\begin{equation}\label{IntCorKernDerxLeft}
\int_0^1 \widetilde{K}_h(x,u)\;du 
= -\frac{g^2(x)}{h^{2}} K\left(\frac{x}{h}\right) \int_0^1 K\left(\frac{x-u}{h}\right)\, du
   + \frac{g(x)}{h} \left[ -K\left(\frac{x-u}{h}\right)\right]_{u=0}^{u=1}
\end{equation}
equals zero for $x\in[0,h]$.
A similar proof might be repeated for $x\in[1-h,1]$.
Finally, equality (\ref{IntCorKern}) holds true for all $x\in(h,1-h)$ too, since
$g(x)\equiv 1$ over this interval.
\medskip

\noindent In what follows, we use more general formulas (\ref{estimator})--(\ref{Gcorrector}) 
instead of (\ref{CorKernLeftG}), that is 
\begin{equation}\label{GcorrectorR}
K_h(x,t)=h^{-1}K((x-t)/h)\,g(x)\,,\quad 
g(x) = \left( \int_{(x-1)/h}^{x/h} K(t)\, dt \right)^{-1},
\quad x\in [0,1]\,.
\end{equation}
Therefore, as follows from (\ref{defKtilde}), (\ref{GcorrectorR})
for any $x\in [0,1]$,
\begin{eqnarray}\label{CorKernLeftG0}
\widetilde{K}_h(x,u)&=& h^{-1}K((x-u)/h)g'(x) + g(x)h^{-2}K'((x-u)/h)\\
&=& \frac{g(x)}{h} \left( \frac{1}{h} K'\left(\frac{x-u}{h}\right) 
 +K\left(\frac{x-u}{h}\right) \left( K_h(x,1) -K_h(x,0) \right) \right).
\label{CorKernLeftG1}
\end{eqnarray}
\medskip

\noindent
The following Lemma proves Lipschitz-like constants in (\ref{eq:AddAssumptions22}) 
and (\ref{eq:LK2primHb})--(\ref{eq:LK2primHa}).

\begin{lemma}\label{LipConstsLem}
Let kernel $K_h$ defined in (\ref{estimator})--(\ref{Gcorrector}) meets the assumptions B1--B3,
and the bandwidth $h\in(0,1/2)$.
Let $\widetilde K_n$ be defined by~(\ref{defKtilde}).
Then the following upper bounds hold true:
\begin{equation}\label{eq:AddAssmptns20L}
\left| \widetilde{K}_h(x,u)\right| \leq 
g_{\max} h^{-2}\left( L_{K} + g_{\max}K_{\max}^2\right) \,,
\end{equation} 
\begin{equation}\label{eq:AddAssmptns22L}
\left| \frac{\partial}{\partial u}\,\widetilde{K}_h(x,u)\right| \leq 
g_{\max} h^{-3} L_{\widetilde{K}} 
\,,\quad
\left| \frac{\partial^2}{\partial u^2}\,\widetilde{K}_h(x,u)\right| \leq 
g_{\max} h^{-4} L_{\widetilde{K}'} \,,
\end{equation} 
where $L_{\widetilde{K}}=L_{K'} +L_{K}\,g_{\max}K_{\max}$
 and $L_{\widetilde{K}'}= L_{K''} +L_{K'}\,g_{\max}K_{\max}$\,.
Moreover,
\begin{equation}\label{eq:LipConst3}
\left|\frac{\partial^3}{\partial x^3}\,K_h(x,u) \right|
 \leq g_{\max} L_{\widetilde{K}''} h^{-4}
\end{equation}
where 
\begin{eqnarray}\label{CoLipTil2pr}
L_{\widetilde{K}''} &=& 
 g_{\max} \left[ L_{K''} + 3L_{K'} K_{\max}g_{\max}  + 3L_Kg_{\max}K_{\max}^2\left(1+3g_{\max}\right) 
\right.\\ \label{CoLipTil2pr1}
&&\left.+\, (L_K^2 +2g_{\max}^2K_{\max}^4)(1+2g_{\max})
\right]\,.
\end{eqnarray}
\end{lemma}
\medskip

\noindent\textbf{Proof of Lemma \ref{LipConstsLem}.}
The upper bound (\ref{eq:AddAssmptns20L}) follows directly from (\ref{CorKernLeftG0})--(\ref{CorKernLeftG1}). 
Furthermore,
taking (\ref{CorKernLeftG0}), (\ref{CorKernLeftG1}) into account, 
one easily may come to (\ref{eq:AddAssumptions22}) since
\begin{equation}\label{eq:AddAssmptns22Pr}
\left| \frac{\partial}{\partial u}\,\widetilde{K}_h(x,u)\right| \leq 
g_{\max} h^{-3} \left( L_{K'} +L_{K}\,g_{\max}K_{\max} \right) = g_{\max} h^{-3} L_{\widetilde{K}} \,,
\end{equation} 
and, similarly,
\begin{equation}\label{eq:AddAssmptns22Pr2}
\left| \frac{\partial^2}{\partial u^2}\,\widetilde{K}_h(x,u)\right| \leq 
g_{\max} h^{-4} \left( L_{K''} +L_{K'}\,g_{\max}K_{\max} \right) = g_{\max} h^{-4} L_{\widetilde{K}'} \,.
\end{equation} 
Moreover,
one may continue calculation of further derivatives from 
(\ref{defKtilde})--(\ref{CorKernLeftG1}) as follows:
\begin{eqnarray}\label{DerivKerCCpr}
 \frac{\partial^2}{\partial x^2}\,K_h(x,u)
&=&  g'(x)h^{-1} \left( h^{-1} K'\left(\frac{x-u}{h}\right) 
    \right.
 \\  \label{DerivKerCC1pr}
&& \phantom{g'(x)h^{-1} \left( \right)}
  \left. +\,K\left(\frac{x-u}{h}\right) \left( K_h(x,1) -K_h(x,0) \right) \right)
 \\ \label{DerivKerCC11pr}
&& +\, g(x)h^{-1} \left( h^{-2} K''\left(\frac{x-u}{h}\right) 
    \right.
 \\ \label{DerivKerCC2pr}
&& \phantom{g(x)h^{-1} \left( \right)}
\left.
 +\,h^{-1} K'\left(\frac{x-u}{h}\right) \left( K_h(x,1) -K_h(x,0) \right)
    \right.
  \\ \label{DerivKerCC21pr}
&& \phantom{g(x)h^{-1} \left( \right)}
 \left. +\, K\left(\frac{x-u}{h}\right) \left( \frac{\partial}{\partial x}\,K_h(x,1) -\frac{\partial}{\partial x}\,K_h(x,0) \right)
  \right)
\end{eqnarray}
and
\begin{eqnarray}\label{DerivKerCCpr2}
 \frac{\partial^3}{\partial x^3}\,K_h(x,u)
&=&  g''(x)h^{-1} \left( h^{-1} K'\left(\frac{x-u}{h}\right) 
    \right.
 \\  \label{DerivKerCC1pr2}
&& \phantom{g'(x)h^{-1} \left( \right)}
  \left. +\,K\left(\frac{x-u}{h}\right) \left( K_h(x,1) -K_h(x,0) \right) \right)
 \\ \label{DerivKerCC11pr2}
&& +\, 2g'(x)h^{-1} \left( h^{-2} K''\left(\frac{x-u}{h}\right) 
    \right.
 \\ \label{DerivKerCC2pr2}
&& \phantom{2g'(x)h^{-1} \left( \right)}
\left.
 +\,h^{-1} K'\left(\frac{x-u}{h}\right) \left( K_h(x,1) -K_h(x,0) \right)
    \right.
  \\ \label{DerivKerCC21pr2}
&& \phantom{2g'(x)h^{-1} \left( \right)}
 \left. +\, K\left(\frac{x-u}{h}\right) \left( \frac{\partial}{\partial x}\,K_h(x,1) -\frac{\partial}{\partial x}\,K_h(x,0) \right)
  \right)
 \\ \label{DerivKerCC11pr3}
&& +\, g(x)h^{-1} \left( h^{-3} K'''\left(\frac{x-u}{h}\right) 
    \right.
 \\ \label{DerivKerCC2pr31}
&& \phantom{g(x)h^{-1} \left( \right)}
 \left.
  +\, h^{-2} K''\left(\frac{x-u}{h}\right) \left( K_h(x,1) -K_h(x,0) \right)
    \right.
 \\ \label{DerivKerCC2pr3}
&& \phantom{g(x)h^{-1} \left( \right)}
 \left.
 +\,2h^{-1} K'\left(\frac{x-u}{h}\right) \left( \frac{\partial}{\partial x}\,K_h(x,1) -\frac{\partial}{\partial x}\,K_h(x,0) \right)
    \right.
  \\ \label{DerivKerCC21pr3}
&& \phantom{g(x)h^{-1} \left( \right)}
 \left. +\, K\left(\frac{x-u}{h}\right) \left( \frac{\partial^2}{\partial x^2}\,K_h(x,1) -\frac{\partial^2}{\partial x^2}\,K_h(x,0) \right)
  \right).
\end{eqnarray}
Moreover, from (\ref{GcorrectorR}) the derivatives follow
\begin{eqnarray}\label{GcorrPrim}
g'(x) &=& g^2(x) h^{-1} \left( K\left(\frac{x-1}{h}\right) -K\left(\,\frac{x}{h}\,\right) \, \right)\,,
\\ \label{GcorrPrim2}
g''(x) &=& 2g(x)g'(x) h^{-1} \left( K\left(\frac{x-1}{h}\right) -K\left(\,\frac{x}{h}\,\right) \, \right)
\\ \label{GcorrPrim21}
&& +\, g^2(x) h^{-2} \left( K' \left(\frac{x-1}{h}\right) -K' \left(\,\frac{x}{h}\,\right) \, \right)\,,
\end{eqnarray}
therefore,
\begin{eqnarray}\label{GcorrPrimUB}
|g'(x)| &\leq& g_{\max}^2 K_{\max} h^{-1}\,,
\\ \label{GcorrPrim2UB}
|g''(x)| &\leq& g_{\max}^2 \left( L_K + 2g_{\max} K_{\max}^2\right) h^{-2} \,.
\end{eqnarray}
Finally, using the bounds (\ref{GcorrPrimUB})--(\ref{GcorrPrim2UB}) in (\ref{DerivKerCCpr2})--(\ref{DerivKerCC21pr3})
and the definition of $\widetilde{K}_h$\, (\ref{GcorrectorR})
 we arrive at the bound (\ref{eq:LK2primHb})--(\ref{eq:LK2primHa}):
\begin{eqnarray}\label{eq:3rdDerivUBL}
h^4 \left|\frac{\partial^3}{\partial x^3}\,K_h(x,u) \right|
&\leq&  g_{\max}^2 \left( L_K + 2g_{\max} K_{\max}^2\right) \left( L_K + g_{\max} K_{\max}^2
 \right)
 \\ \label{DerivKerCC11pr2UBL}
&& +\, 2g_{\max}^2 K_{\max}  \left( L_{K'}^{} + L_{K} g_{\max} K_{\max} 
  \right.
  \\ \label{DerivKerCC21pr2UBL}
&& \phantom{2g'(x)h^{-1}(a) \left( \right)}
  +\, K_{\max} h^2 \max_{x,u}   | \widetilde{K}_h(x,u) | \,)
  \\ \label{DerivKerCC11pr3UBL}
&& +\, g_{\max}  (L_{K''}  + L_{K'} g_{\max} K_{\max} 
+2L_K h^2 \max_{x,u}  | \widetilde{K}_h(x,u) |   
 \\ \label{DerivKerCC21pr3UBL}
&& \phantom{g(x) (aaa)}
+\, K_{\max} h^3 \max_{x,u} | \partial\,\widetilde{K}_h(x,u)/{\partial x}|   \,)
 \\ \label{eq:3rdDeriv1UBL}
 &\leq& g_{\max}   \left[ L_{K''} + 3L_{K'} g_{\max} K_{\max} \right.
   \\ \label{eq:3rdDeriv1UBL1} 
 && \phantom{g_{\max}   \left[\right.} +\, 3L_K g_{\max} \left(L_K+K_{\max}^2g_{\max} \right) 
 \\ \label{eq:3rdDeriv1}		&&\left.
 					\phantom{g_{\max}  \left[\right.}
					 +\, K_{\max} g_{\max}^2 \left( 3K_{\max} L_K +2K_{\max}^3 g_{\max}^2 \right)
 				\right]
 \\ \label{eq:3rdDeriv2}
 &=& g_{\max} \left[ L_{K''} + 3L_{K'} K_{\max}g_{\max}  + 3L_Kg_{\max}K_{\max}^2\left(1+3g_{\max}\right) 
\right.
 \\ \label{eq:3rdDeriv21}	&&\left.
 					\phantom{g_{\max}   \left[\right.}
+\, (L_K^2 +2g_{\max}^2K_{\max}^4)(1+2g_{\max})
 					\right].
\end{eqnarray}
Here we applied the upper bound (\ref{eq:AddAssmptns20L})
as well as the one, followed from (\ref{CorKernLeftG0})--(\ref{CorKernLeftG1}) 
and (\ref{DerivKerCCpr})--(\ref{DerivKerCC21pr}):
\begin{eqnarray}\label{partial2UB}
h^3\max_{x,u} \left| \frac{\partial}{\partial x}\,\widetilde{K}_h(x,u) \right| 
&\leq& g_{\max}^2 K_{\max} 
   \left(  L_K + g_{\max} K_{\max}^2  \right)
 \\ \label{DerivKerCC11prP}
&& +\, g_{\max} h^{-3} \left( L_{K'} + L_K g_{\max} K_{\max} 
    \right.
 \\ \label{DerivKerCC2prP}
&& \phantom{g(x)h^{-1} \left( a\right)}
 \left. +\,K_{\max} g_{\max} \left(L_K +K_{\max}^2 g_{\max}\right)
  \right).
\end{eqnarray}
Lemma~\ref{LipConstsLem} is proved.
\CQFD

\subsection{Auxiliary lemmas.~I}
\label{subsecAuxLemsI}

The following results are proved here for the sake of completeness.

\begin{lemma}\label{CXlemma}
Let function $f:[0,1]\to\R$ 
meets the assumption A1 and sequence $(X_i)_{i=1,\dots,N}$ be
obtained from an independent
sample with p.d.f. $f(x)/C_f$ by increase ordering (\ref{Xorder}),
where $C_f$ is defined by (\ref{eq:Cf}).
Denote $X_0=0$ and $X_{N+1}=1$.
Then for any finite constant $C_X > 4C_f/f_{\min}$
there exist almost surely finite
number $N_0=N_0(\omega)$ such that
\begin{equation}\label{LemmaX}
\max_{i=1,\dots,N+1} \Delta{X_i}  \leq C_X \frac{\log{N}}{N}
\quad \forall \;N\geq N_0
\end{equation}
with probability 1.
For instance, one may fix constant $C_X$ as follows:
\begin{equation}\label{eq:CXvalue}
C_X= 5f_{\max}/f_{\min}\,.
\end{equation}
\end{lemma}
\medskip

\noindent \textbf{Proof of Lemma~\ref{CXlemma}.}
Introduce a uniform partition of the interval $[0,1]$ onto $m_N^{}$
subintervals $\Delta_k$ with equal Lebesgue measures
\begin{equation}\label{eq:DeltaSize}
    \ell(\Delta_k) \triangleq 1/m_N^{} \leq C_X \log{N}/(2N)\,,\quad
     k=1,\dots,m_N^{}\,,
\end{equation}
where size of partition
\begin{eqnarray}\label{eq:mNnumber}
m_N^{} &\triangleq& \min\{\mathrm{integer}\; m\,:\; m\geq{2N}/(C_X \log{N}) \}
\\ \label{eq:mNnumber1}
&\leq& 1+ \frac{2N}{C_X \log{N}}  \leq \frac{(2+\varepsilon)N}{C_X \log{N}}
\end{eqnarray}
for an arbitrary $\varepsilon>0$ and for any sufficiently large $N$.
Hence, the event
\begin{eqnarray}\label{eq:ANevent}
 A_N^{} &\triangleq& \{\omega\,:\; \max_{i=1,\dots,N+1}\Delta{X_i}\leq C_X \log{N}/N \}
  \\ \label{eq:ANevent1}
   &\supseteq& \bigcap_{k=1}^{m_N^{}}
   \left[\bigcup_{i=1}^N \left\{ X_i \in \Delta_k \right\}
    \right] \,.
\end{eqnarray}
Basing on Borel--Cantelli lemma we prove that the complementary event
$A_N^{c}\triangleq\Omega\setminus A_N^{}$
may occur only finite number of times (with probability 1).
Evidently,
\begin{eqnarray}\label{ProbAN}
 P\left(A_N^{c}\right) &\leq & \sum_{k=1}^{m_N^{}} P\left(
   \bigcap_{i=1}^N \left\{ X_i \notin \Delta_k \right\} \right)
  \\ \label{ProbAN1}
  &=& \sum_{k=1}^{m_N^{}} \prod_{i=1}^N \left( 1- \int_{\Delta_k} C_f^{-1}f(u)\,du \right)
    \\ \label{ProbAN2}
   &\leq& m_N^{} \left( 1- \frac{f_{\min}}{C_f^{}}\cdot \ell(\Delta_1) \right)^N
    \\ \label{ProbAN3}
   &\leq& \frac{(2+\varepsilon)N}{C_X \log{N}}\,
        \exp\left\{- \frac{f_{\min} C_X}{(2+\varepsilon)C_f^{}}\cdot \log{N} \right\}
    \\ \label{ProbAN4}
   &=& O\left(N^{1-{f_{\min} C_X}/((2+\varepsilon)C_f^{})}\right)\,.
\end{eqnarray}
Hence, condition $C_X>4C_f/f_{\min}$ implies the existence of positive $\varepsilon$ 
ensuring the convergence of series
\begin{equation}\label{eq:BorelCantelli}
    \sum_{N=1}^\infty P\left(A_N^{c}\right) <\infty\,,
\end{equation}
and the Borel--Cantelli lemma applies.
Note, that events $\bigcap_{i=1}^N \left\{ X_i \notin \Delta_k \right\}$
do not depend on renumbering of $(X_i)_{i=1,\dots,N}$ which lead to (\ref{ProbAN1})
from (\ref{ProbAN});
moreover, we have used both definition (\ref{eq:mNnumber}) and
inequality $1-x\leq e^{-x}$ there in (\ref{ProbAN2})--(\ref{ProbAN3}).
Lemma~\ref{CXlemma} is proved.
\CQFD
\medskip

\begin{lemma}\label{AuxPartiLemma}
Let random sample $\{(X_i,Y_i)\, |\,\,i=1,\dots,N\}$ be
defined as in Section~\ref{subsecLPprobl}.
Let sequence $\delta_x=\delta_x(N)$ be positive,  and for some $\varepsilon>0$
\begin{equation}\label{eq:liminfN1eps}
\liminf_{N\to\infty}N^{1-\varepsilon}\delta_x>0\,. 
\end{equation}
Define
\begin{equation}\label{eq:mdeltaDef}
m_{\delta}^{} \triangleq \min\{\mathrm{integer}\; m\,:\; m\geq \delta_x^{-1} \}
\end{equation}
and assume a positive sequence $\delta_y=\delta_y(N)<f_{\min}$
meeting for all sufficiently large $N$ the inequality
\begin{equation}\label{eq:deltayDef}
\delta_y^{}\geq \kappa\, m_{\delta}^{}\frac{\log N}{N}\,,\quad \mathrm{with}\quad\kappa> \frac{(2-\varepsilon)C_f}{f_{\min}}\,.
\end{equation} 
Then, under the assumptions of Lemma~\ref{CXlemma}, with probability 1, there exists finite number $N_6(\omega)$ such that for any $N\geq N_6(\omega)$
there is such a subset of points $\left\{\left(X_{i_k},Y_{i_k}\right),\,
k=1,\dots,m_{\delta}^{}\right\}$\, in the sample $\left\{\left(X_{i},Y_{i}\right),\right.$
$\left. i=1,\dots,N\right\}$, that the following inequalities hold:
\begin{equation}\label{eq:AuxPartiLemma}
(k-1)/m_{\delta}^{}\leq X_{i_k}< k/m_{\delta}^{}\,,\quad
f(X_{i_k})-\delta_y^{}\leq Y_{i_k}\leq f(X_{i_k})\,.
\end{equation}
\end{lemma}

 \noindent \textbf{Proof of Lemma~\ref{AuxPartiLemma}.}
 It is similar to that of Lemma~\ref{CXlemma}.
Introduce an equidistant partition of the interval $[0,1]$ onto 
subintervals $[(k-1)/m_{\delta}^{},k/m_{\delta}^{}]$, $k=1,\dots,m_{\delta}^{}$\,.
Moreover, introduce the related subsets in $\R^2$ 
\begin{equation}\label{eq:DeltaH}
    \Delta_k\triangleq \{ (u,v) : (k-1)/m_{\delta}^{}\leq u\leq k/m_{\delta}^{}\,,\; f(u)-\delta_y^{}\leq v\leq f(u)\}
     \,,\quad
     k=1,\dots,m_{\delta}^{}\,.
\end{equation}
Hence, the event
\begin{eqnarray}\label{eq:ANeventH}
 A_N^{} &\triangleq& \{\omega\,:\; \forall\,k=1,\dots,m_{\delta}^{}\,\,\exists\, i=1,\dots,N : \left( X_{i},Y_{i}\right)\in\Delta_k \}
  \\ \label{eq:ANeventH1}
   &=& \bigcap_{k=1}^{m_{\delta}^{}}
   \left[\bigcup_{i=1}^N \left\{ \left( X_{i},Y_{i}\right)\in\Delta_k \right\}
    \right] \,.
\end{eqnarray}
Basing on Borel--Cantelli lemma we prove that the complementary event
$A_N^{c}\triangleq\Omega\setminus A_N^{}$
may occur only finite number of times (with probability 1).
Evidently,
\begin{eqnarray}\label{ProbANh}
 P\left(A_N^{c}\right) &\leq & \sum_{k=1}^{m_{\delta}^{}} P\left(
   \bigcap_{i=1}^N \left\{ \left( X_{i},Y_{i}\right) \notin \Delta_k \right\} \right)
  \\ \label{ProbANh1}
  &=& \sum_{k=1}^{m_{\delta}^{}} \prod_{i=1}^N \left( 1- \int_{\Delta_k} C_f^{-1}f(u)\,du\,dv \right)
    \\ \label{ProbANh2}
   &\leq& m_{\delta}^{} \left( 1- \frac{f_{\min}}{C_f^{}}\cdot \frac{\delta_y^{}}{m_{\delta}^{}} \right)^N
    \\ \label{ProbANh3}
   &\leq& \left(1+\delta_x^{-1}\right)\,
        \exp\left\{- \frac{f_{\min} \kappa}{C_f^{}}\cdot \log{N} \right\}
    \\ \label{ProbANh4}
   &=&  O\left(N^{1-\varepsilon-{f_{\min} \kappa}/C_f^{}}\right)\,.
\end{eqnarray}
Hence, condition $\kappa>(2-\varepsilon)C_f/f_{\min}$ implies 
\begin{equation}\label{eq:BorCanti}
    \sum_{N=1}^\infty P\left(A_N^{c}\right) <\infty\,,
\end{equation}
and one may apply  Borel--Cantelli lemma. 
Lemma~\ref{AuxPartiLemma} is proved.
\CQFD
\medskip

\begin{Coro}\label{CorAuxPartiLemma}
Let $\delta_x$ and $\delta_y$ meet the conditions of Lemma~\ref{AuxPartiLemma}. Then, with probability~1,  
for any $N\geq N_6(\omega)$ and
any $x\in[0,1]$ there exists integer $i_k\in\{1,\dots,N\}$
such that $|x-X_{i_k}|\leq \delta_x$ and $ f(X_{i_k})-\delta_y^{}\leq Y_{i_k}\leq f(X_{i_k})$\,.
\end{Coro}

\medskip

\begin{lemma}\label{AuxLemma}
Let function $g:[0,\Delta]\to\R$ be twice continuous differentiable, $\Delta>0$.
Then
\begin{equation}\label{eq:AuxLemma}
    \max_{x\in[0,\Delta]} |g(x)| \leq \max\{ |g(0)|, |g(\Delta)| \}
     +\frac{\Delta^2}{8} \max_{x\in[0,\Delta]} |g^{\,\prime\prime}(x)|
     \,.
\end{equation}
\end{lemma}

 \noindent \textbf{Proof of Lemma~\ref{AuxLemma}.}
Denote $\bar{g}_b^{}=\max\{ |g(0)|, |g(\Delta)| \}$. It suffices to prove
the case where a point  $x_1\in(0,\Delta)$ exists with
\begin{equation}\label{eq:Bar}
    |g(x_1)| =\max_{x\in[0,\Delta]} |g(x)| > \bar{g}_b^{}\,.
\end{equation}
Then  $g^{\,\prime}(x_1)=0$, and for any $x\in[0,\Delta]$
\begin{equation}\label{eq:IntRepres}
g(x_1)=g(x) -\int_{x_1}^x dt \int_{x_1}^t g^{\,\prime\prime}(u)\,du\,.
\end{equation}
Therefore, putting $x=\Delta$ one obtains from (\ref{eq:IntRepres})
\begin{equation}\label{eq:IntRepresR}
|g(x_1)|\leq |g(\Delta)|
 +\int_{x_1}^{\Delta} dt \int_{x_1}^t |g^{\,\prime\prime}(u)|\,du
\leq \bar{g}_b^{} + \frac{(\Delta-x_1)^2}{2}\max_{x\in[0,\Delta]} |g^{\,\prime\prime}(x)|
 \,.
\end{equation}
Similarly, fixing  $x=0$ there in (\ref{eq:IntRepres}) leads to
\begin{equation}\label{eq:IntRepresL}
|g(x_1)|\leq |g(0)|
 +\int_{0}^{x_1} dt \int_{x_1}^t |g^{\,\prime\prime}(u)|\,du
\leq \bar{g}_b^{} + \frac{x_1^2}{2}\max_{x\in[0,\Delta]} |g^{\,\prime\prime}(x)|
 \,.
\end{equation}
Thus, combining (\ref{eq:IntRepresR}) and (\ref{eq:IntRepresL}) we arrive at
\begin{equation}\label{eq:IntRepresMax}
|g(x_1)| \leq \bar{g}_b^{}
 + \frac{1}{2} \min\{(\Delta-x_1)^2 , x_1^2 \}
  \max_{x\in[0,\Delta]}|g^{\,\prime\prime}(x)|
 \,.
\end{equation}
Since
\begin{equation}\label{eq:MaxMin}
\max_{x\in[0,\Delta]} \min\{(\Delta-x)^2 , x^2 \} = \frac{\Delta^2}{4}\,,
\end{equation}
the desired inequality (\ref{eq:AuxLemma}) follows immediately from (\ref{eq:Bar}),
(\ref{eq:IntRepresMax})--(\ref{eq:MaxMin}).
\CQFD

\subsection{Auxiliary lemmas.~II}
\label{subsecAuxLemsII}

Lemma~\ref{SumDX3lemma} states the results announced in the Remark~\ref{rem:lognu}, Subsection~\ref{subsec:UB}.
\begin{lemma}\label{SumDX3lemma}
Let numbers $h_N$ form a positive sequence, non-increasing for $N\geq N_1$ and 
meeting condition
\begin{equation}\label{eq:hNN-1}
\frac{h_{N-1}}{h_N}\leq 1+\frac{\kappa}{N}
\quad\forall\,N\geq N_1
\end{equation}
with finite, positive constants $\kappa$ and $N_1$ and such that
\begin{equation}
\label{proplog}
\lim_{N\to\infty} \frac{\log N}{N h_N} =0
\,.
\end{equation}
Then, under the assumptions of Lemma~\ref{CXlemma}, as $N\to\infty$, for an arbitrary $\nu>1$ and for any $x\in[0,1]$
\begin{equation}\label{eq:SumDX3}
S_N\triangleq
\sum_{i=1}^{N+1} (\Delta X_i)^3\,\mathbf{1}\{|x-X_i|\leq 2h_N\}
=o\left(\frac{h_N\log^\nu{N}}{N^2}\right)\quad \mathrm{a.s.}
\end{equation}
where $o(\cdot)$
does not depend on $x$.
\end{lemma}
\smallskip

 \noindent \textbf{Proof of Lemma~\ref{SumDX3lemma}.}
\par\medskip\noindent
1. Remind that the sequence of random points $(X_i)$ is obtained from that of
i.i.d. with the p.d.f. $f(\cdot)/C_f$ for $X_i$ by their increase ordering.
Furthermore, $\Delta X_i\triangleq X_i-X_{i-1}$, $X_0=0$, and $X_N\equiv 1$.
Introduce $\sigma$-algebras $\mathcal{F}_N\triangleq\sigma\{X_1,\dots,X_N\}$.
Thus, $(S_N,\mathcal{F}_N)$ is a non-negative stochastic sequence. 
Let us denote $X$ the new point (hence, independent of $\mathcal{F}_N$)
when passing from $S_N$ to $S_{N+1}$. With these notations and due to the evident
inequality 
$$\mathbf{1}\{|x-X_i|\leq 2h_{N+1}\} \leq \mathbf{1}\{|x-X_i|\leq 2h_{N}\}$$
one may write
\begin{eqnarray}\label{eq:CondEspeSN}
\Espe (S_{N+1} | \mathcal{F}_{N}) 
&\leq& \Espe \left\{ \sum_{j=1}^{N+1} \mathbf{1}\{X\in[X_{j-1},X_j)\} 
\cdot \left[ \sum_{i\neq j}^{N+1} (\Delta{X}_i)^3 \mathbf{1}\{|x-X_i|\leq 2h_N\}
\right.\right.
\\ \label{eq:CondEspeSN1}
 &&  
 \phantom{\sum_{j=1}^{N+1} \sum_{j=1}^{N+1}}
 +\left( (X-X_{j-1})^3  \mathbf{1}\{|x-X|\leq 2h_N\}
 \right.
\\ \label{eq:CondEspeSN12}
 && \left.\left. \phantom{\sum_{j=1}^{N+1} \sum_{j=1}^{N+1}}
     \left.
         +(X_{j}-X)^3 \mathbf{1}\{|x-X_j|\leq 2h_N\} \right)
\right] | \mathcal{F}_{N}
\right\}
\\ 
\label{eq:CondEspeSN2}
&\leq& S_N -\sum_{j=1}^{N+1} \mathbf{1}\{|x-X_j|\leq 2h_N\}\,
\Espe\left\{  \mathbf{1}\{X\in[X_{j-1},X_j)\} 
\right.
\\ \label{eq:CondEspeSN3}
 &&  \phantom{S_{N+1}-S_{N}} \left.\cdot \left[ (\Delta{X}_j)^3
 -(X-X_{j-1})^3 -(X_{j}-X)^3
\right] | \mathcal{F}_{N}
\right\}
\\ \label{eq:CondEspeSN4}
 && 
+\,\sum_{j=1}^{N+1} \Espe\left\{ (X-X_{j-1})^3 \mathbf{1}\{X\in[X_{j-1},X_j)\}  \left( \mathbf{1}\{|x-X|\leq 2h_N\} \right.
     \right.
\\ \label{eq:CondEspeSN41}
 && \left. \phantom{S_{N+1}-S_{N}S_{N} (X-X_{j-1})^3 } \left. - \mathbf{1}\{|x-X_j|\leq 2h_N\} \right)
\, | \mathcal{F}_{N} \right\}\,.
\end{eqnarray}
\medskip
\noindent
2. The first need now is to evaluate the conditional expectation in (\ref{eq:CondEspeSN2})--(\ref{eq:CondEspeSN3}).
A simple algebras imply
$$
(\Delta{X}_j)^3  -(X-X_{j-1})^3-(X_{j}-X)^3 
= 3(X(X_j+X_{j-1}) -X^2 -X_j X_{j-1})\Delta{X}_j
$$
which is non-negative for any $X\in[X_{j-1},X_j)$.
Therefore, the bounding from below leads to
\begin{equation}\label{eq:CondEspeDelxxx}
\Espe\left\{  \mathbf{1}\{X\in[X_{j-1},X_j)\} \left[ (\Delta{X}_j)^3
 -(X-X_{j-1})^3-(X_{j}-X)^3
\right] | \mathcal{F}_{N}
\right\}
\end{equation}
\begin{eqnarray}\label{eq:CondEspeDelxxx1}
&=& 3\Delta{X}_j \int_{X_{j-1}}^{X_j} \frac{f(x)}{C_f}\,(x(X_j+X_{j-1}) -x^2 -X_j X_{j-1})\,dx
\\ \label{eq:CondEspeDelxxx2}
&\geq&  \frac{f_{\min}}{2C_f} \left(\Delta{X}_j\right)^4.
\end{eqnarray}
Substituting to (\ref{eq:CondEspeSN2})--(\ref{eq:CondEspeSN3}) and
applying Iensen's inequality for the convex function $\psi(s)\triangleq s^{4/3}$, 
\begin{equation}\label{eq:Iensen}
\left(\frac{S_N}{\#\{|x-X_j|\leq 2h_N\}}\right)^{4/3} \leq
\frac{\sum_{j=1}^{N+1} \mathbf{1}\{|x-X_j|\leq 2h_N\}\, \left(\Delta{X}_j\right)^4}
      {\#\{|x-X_j|\leq 2h_N\}}\,,
\end{equation}
 lead to 
\begin{equation}\label{eq:CondEspeSNew1}
\Espe (S_{N} | \mathcal{F}_{N-1}) 
\leq S_{N-1} - \frac{f_{\min}}{2C_f}\,\frac{S_{N-1}^{4/3}}{(Nq_N)^{1/3}}
+ r_N
\end{equation}
where
\begin{equation}\label{eq:EventNUmber}
q_N \triangleq \frac{1}{N}\sum_{j=1}^{N} \mathbf{1}\{|x-X_j|\leq 2h_{N-1}\} 
=O(h_{N-1})\,,
\end{equation}
and $r_N$ denotes the related sum in (\ref{eq:CondEspeSN4})--(\ref{eq:CondEspeSN41}), that is
\begin{eqnarray}\label{eq:DEFrN}
r_N &\triangleq & 
\sum_{j=1}^{N} \Espe\left\{ (X-X_{j-1})^3 \mathbf{1}\{X\in[X_{j-1},X_j)\}  \left( \mathbf{1}\{|x-X|\leq 2h_{N-1}\} \right.
     \right.
\\ \label{eq:DEFrN1}
 && \left. \phantom{S_{N+1}-S_{N}S_{N} (X-X_{j-1})^3 } \left. - \mathbf{1}\{|x-X_j|\leq 2h_{N-1}\} \right)
\, | \mathcal{F}_{N-1} \right\}\,.
\end{eqnarray}
Note, that one may define $0/0\triangleq 0$ to treat the case of zero denominators there in (\ref{eq:Iensen}), for instance.

The bound $O(h_N)$ for $q_N$ stated in (\ref{eq:EventNUmber}) is proved below in Lemma~\ref{qNbounding}.
In order to bound $r_N$ from above one may easily see that the difference between the two indicators in (\ref{eq:DEFrN})--(\ref{eq:DEFrN1}) is positive iff the first of them equals $1$
while the second does $0$. 
Due to Lemma~\ref{CXlemma} and the property~(\ref{proplog}), i.e.  
$\log N/(Nh_N)\to 0$,
this may almost surely arise only for the  following event (for any sufficiently large $N$): 
$x-2h_{N-1}\leq X_{j-1}<X\leq x+2h_{N-1}<X_j$\,.
Given a sequence $(X_i)$, this event arises only for one $j$, say $j=j_{0}^{}$,
which depends on $x$\,.
Thus, 
\begin{eqnarray}\label{eq:DEFrN0}
r_N &\leq& \int_{X_{j_{0}^{}-1}}^{X_{j_{0}^{}}} (u-X_{j_{0}^{}-1})^3 \frac{f(u)}{C_f}\,du
\leq \frac{f_{\max}}{C_f} \max_{i=1,\dots,N+1} (X_i-X_{i-1})^4
\\ \label{eq:DEFrN01}
&=&O\left(\left( \log N/N\right)^4 \right),
\end{eqnarray}
with non-random $O(\cdot)$ being independent of $x$, from Lemma~\ref{CXlemma}.
\medskip

\noindent
3. The next step is to come from the nonlinear inequality (\ref{eq:CondEspeSNew1}) to a linear one.
The convexity of function $\psi(s)=s^{4/3}$ gives, for an arbitrary $a_N>0$, the lower bound as follows:
$$
\psi(S_N)\geq \psi(a_N) +\psi'(a_N)(S_N-a_N) = \frac{4}{3} a_N^{1/3} S_N -\frac{1}{3} a_N^{4/3}\,.
$$
Thus, inequality (\ref{eq:CondEspeSNew1}) and the choice
$a_{N-1}\triangleq a^3 q_N/N^2$ with 
 $a>0$
lead to
\begin{eqnarray}\label{eq:CondEspeSNews}
\Espe (S_{N} | \mathcal{F}_{N-1}) 
&\leq& S_{N-1} - \frac{2f_{\min}}{3C_f} \left(\frac{a_{N-1}}{Nq_N}\right)^{1/3} S_{N-1}
 + \frac{f_{\min}}{6C_f} \left(\frac{a_{N-1}^4}{Nq_N}\right)^{1/3} 
 + r_N
\\ \label{eq:CondEspeSNews1}
&\leq& S_{N-1} -  \frac{\mu}{N}  S_{N-1} +\frac{\mu a^3}{4N^3} q_N
+ r_N
\end{eqnarray}
where
\begin{equation}\label{eq:MuD}
\mu\triangleq\frac{2af_{\min}}{3C_f}>2+\kappa
\end{equation}
for sufficiently large $a$. 
\medskip

\noindent
4. Finally, using relations (\ref{eq:EventNUmber})--(\ref{eq:DEFrN1}) in 
(\ref{eq:CondEspeSNews})--(\ref{eq:CondEspeSNews1})
 and applying Lemma~\ref{MartngLemma} we arrive at the result of
Lemma~\ref{SumDX3lemma}.
\CQFD

\begin{lemma}\label{MartngLemma}
Let $(w_N,\mathcal{F}_N)$ be non-negative stochastic sequence meeting the inequality
\begin{equation}\label{eq:MartnglIneq0}
\Espe (w_N | \mathcal{F}_{N-1}) \leq \left(1-\frac{\mu}{N}\right)w_{N-1} +\frac{d_N h_N}{N^{1+p}}
\quad \mathrm{a.s.} \quad\forall\,N\geq N_1
\end{equation}
where $\mu>p+\kappa$, $h_N$ and $\kappa$ meet conditions of
Lemma~\ref{SumDX3lemma},
$d_N$ is $\mathcal{F}_{N-1}$-measurable, non-negative and bounded a.s., 
and $N_1<\infty$.
Then, as $N\to\infty$,  for any $\nu>1$
\begin{equation}\label{eq:MartnglIneq}
w_N =o\left( \frac{\log^\nu{N}}{N^{p}} h_N \right)
\quad \mathrm{a.s.}
\end{equation}
\end{lemma}

\noindent \textbf{Proof of Lemma~\ref{MartngLemma}.}
Introduce
\begin{equation}\label{eq:vNintro}
v_N \triangleq \frac{N^{p}}{h_N\log^\nu{N}} \,w_N
\quad \mathrm{a.s.}
\end{equation}
The inequalities $\log N>\log(N-1)$, $h_{N-1}\geq h_N$, and (\ref{eq:MartnglIneq0}) imply
\begin{eqnarray}\label{eq:MartnglIneq1}
\Espe (v_N | \mathcal{F}_{N-1}) &\leq& \left(1-\frac{\mu}{N}\right)\left(\frac{N}{N-1}\right)^{p} 
 \frac{h_{N-1}}{h_N}\, v_{N-1} +\frac{d_N}{N\log^\nu{N}}
\\ \label{eq:MartnglIneq2}
&\leq& \left(1-\frac{\mu-p-\kappa}{N} +o\left(\frac{1}{N}\right)\right) v_{N-1} + \frac{d}{N\log^\nu{N}}
\end{eqnarray}
Since
\begin{equation}\label{eq:ConvSeries}
\sum^\infty\frac{\mu-p-\kappa}{N}=\infty
\quad\mathrm{and}\quad
\sum^\infty \frac{d_N}{N\log^\nu{N}} <\infty
\quad\mathrm{a.s.}\,,
\end{equation}
one may apply Robbins--Siegmund almost supermartingale convergence theorem 
{\sc Robbins} \& {\sc Siegmund}~\cite{RobSieg}
which implies $v_N\to{0}$ as $N\to\infty$.
Lemma~\ref{MartngLemma} is proved.
\CQFD

\begin{lemma}\label{qNbounding}
Let $h=h_N\to 0$ as $N\to\infty$.  
Then,
under the assumptions of Lemma~\ref{CXlemma} and Lemma~\ref{SumDX3lemma},
the bound (\ref{eq:EventNUmber}) holds true for any $x\in[0,1]$, that is
\begin{equation}\label{eq:EventNUmberPr}
q_N \triangleq \frac{1}{N}\sum_{j=1}^{N} \mathbf{1}\{|x-X_j|\leq 2h_{N-1}\} 
=O(h_N)
\end{equation}
where $O(\cdot)$ does not depend on $x$.
\end{lemma}
\medskip

\noindent \textbf{Proof of Lemma~\ref{qNbounding}.}
Introduce
\begin{equation}\label{eq:zetaIntro}
\zeta_{i} \triangleq \mathbf{1}\{|x-X_i|\leq 2h_{N-1}\} - P\{|x-X_i|\leq 2h_{N-1} \}
\end{equation}
leading to the decomposition
\begin{equation}\label{eq:EventNUmberDecomp}
q_N = P\{|x-X_1|\leq 2h_{N-1} \} + \frac{1}{N}\sum_{j=1}^{N} \zeta_{i}
\,.
\end{equation}
Since $X_i$ are i.i.d. with the bounded p.d.f. $f(\cdot)/C_f$, 
the probability 
\begin{equation}\label{eq:Pro}
P\{|x-X_i|\leq 2h_{N-1} \} \leq \int_{x-2h_{N-1}}^{x+2h_{N-1}}  \frac{f(u)}{C_f}\,du =O(h_{N-1}) 
\end{equation}
with $O(\cdot)$ being independent of $x$ and of $i$.
Furthermore, observe that $|\zeta_{i}|\leq 1$ a.s., and
 \begin{equation}\label{eq:zetaMoms}
\Espe\zeta_{i}=0 \,,\qquad \Espe\zeta_{i}^2 \leq  P\{|x-X_i|\leq 2h_{N-1} \} 
=O(h_{N-1})\,.
\end{equation}
Thus, in order to bound the stochastic term in the right hand side (\ref{eq:EventNUmberDecomp})
one may apply the Bernstein inequality (see, e.g.,
{\sc Birg\'e} \& {\sc Massart}~\cite{BirMas} or {\sc Bosq}~\cite{Bosq},
Theorem 2.6) 
with the standard treatment via Borel--Cantelli lemma
(e.g., as in {\sc Bouchard} {\it et al}~\cite{BGINrep}, Appendix, Lemma~5).
This directly yields
\begin{equation}\label{eq:DecompRes}
q_N =O(h_N) + O\left(\left(\frac{h_N\log N}{N} \right)^{1/2}\right)
=O(h_N)
\quad \mathrm{a.s.}
\end{equation}
Lemma~\ref{qNbounding} is proved.
\CQFD


\end{document}